\documentclass[final,journal,letterpaper,twoside,twocolumn]{IEEEtran}

\ifCLASSINFOpdf
  \usepackage[pdftex]{graphicx}
  \graphicspath{{./figures/}}
  \DeclareGraphicsExtensions{.eps,.pdf,.jpeg,.png,.tif,.jpg}
\else
\fi
\usepackage{amsmath}
\usepackage{algorithm,algorithmic}

\usepackage{booktabs}

\usepackage{xcolor}

\usepackage{array}

\ifCLASSOPTIONcompsoc
  \usepackage[caption=false,font=normalsize,labelfont=sf,textfont=sf]{subfig}
\else
  \usepackage[caption=false,font=footnotesize]{subfig}
\fi

\begin{document}
%
\title{A fast and accurate basis pursuit denoising algorithm with application to super-resolving tomographic SAR}
%
%
%

\author{Yilei~Shi,
        Xiao Xiang~Zhu,~\IEEEmembership{Senior~Member,~IEEE},
        ~Wotao Yin and~Richard~Bamler,~\IEEEmembership{Fellow,~IEEE}
\thanks{This work is supported by the European Research Council (ERC) under the European Union's Horizon 2020 research and innovation programme (grant agreement no. ERC-2016-StG-714087, acronym: So2Sat, www.so2sat.eu), the Helmholtz Association under the framework of the Young Investigators Group ``SiPEO" (VH-NG-1018, www.sipeo.bgu.tum.de), Munich Aerospace e.V. – Fakult{\"a}t f{\"u}r Luft- und Raumfahrt, and the Bavaria California Technology Center (Project: Large-Scale Problems in Earth Observation). The authors thank the Gauss Centre for Supercomputing (GCS) e.V. for funding this project by providing computing time on the GCS Supercomputer SuperMUC at the Leibniz Supercomputing Centre (LRZ). (\textit{Corresponding Author: Xiao Xiang Zhu})}
\thanks{Y.~Shi with the Chair of Remote Sensing Technology (LMF), Technische Universität München (TUM), 80333 Munich, Germany (e-mail: yilei.shi@tum.de)}
\thanks{X.X.~Zhu with the Remote Sensing Technology Institute
(IMF), German Aerospace Center (DLR) and Signal Processing in Earth Observation (SIPEO), Technische Universität München (TUM), 80333 Munich, Germany (e-mail: xiao.zhu@dlr.de)}
\thanks{W. Yin with the Department of Mathematics, University of Califonia, Los Angeles (UCLA), Los Angeles, CA 90095-1555 (e-mail: wotaoyin@math.ucla.edu)}
\thanks{R. Bamler with the Remote Sensing Technology Institute
(IMF), German Aerospace Center (DLR) and the Chair of Remote Sensing Technology (LMF), Technische Universität München (TUM), 80333 Munich, Germany (e-mail: richard.bamler@dlr.de)}}

%
%

\markboth{IEEE TRANSACTIONS ON GEOSCIENCE AND REMOTE SENSING, in press}%
{Y. Shi \MakeLowercase{\textit{et al.}}: A fast and accurate basis pursuit denoising algorithm with application to super-resolving tomographic SAR}
%



\maketitle

\begin{abstract}\\

\textcolor{blue}{\textit{This is the pre-print version, to read the final version please go to IEEE Transactions on Geoscience and Remote Sensing on IEEE XPlore.}}\\

$L_1$ regularization is used for finding sparse solutions to an underdetermined linear system. As sparse signals are widely expected in remote sensing, this type of regularization scheme and its extensions have been widely employed in many remote sensing problems, such as image fusion, target detection, image super-resolution, and others and have led to promising results. However, solving such sparse reconstruction problems is computationally expensive and has limitations in its practical use. In this paper, we proposed a novel efficient algorithm for solving the complex-valued $L_1$ regularized least squares problem. Taking the high-dimensional tomographic synthetic aperture radar (TomoSAR) as a practical example, we carried out extensive experiments, both with simulation data and real data, to demonstrate that the proposed approach can retain the accuracy of second order methods while dramatically speeding up the processing by one or two orders. Although we have chosen TomoSAR as the example, the proposed method can be generally applied to any spectral estimation problems.

\end{abstract}


\section{Introduction}
In this paper, we focus on the so-called $L_1$ regularized least squares (L1LS) minimization problem of the following form:

\begin{equation}
\min_{\bf{x}}  \Vert \bf{Ax}-\bf{b} \Vert^2_2 + \lambda \Vert \bf{x} \Vert_1
\label{equ:opt_l1lsp_org}
\end{equation}

It is an unconstrained convex optimization problem with a non-differentiable objective function due to the presence of the $L_1$ term. This L1LS problem is also known as the basis pursuit denoising approach (BPDN) or least absolute shrinkage and selection operator (LASSO). It promotes sparse solutions, which are commonly desired in many applications in computer vision, machine learning or other fields. Sparsity is also widely exploited in remote sensing. For example, for multi- and hyperspectral sensors, it is used for pan-sharpening \cite{bib:Zhu2013} and spectral unmixing \cite{bib:Bieniarz2015}. It is also used for spectral estimation in tomographic SAR \cite{bib:Zhu2010a}\cite{bib:Budillon2011}, rational polynomial coefficients' estimation of rational function model for photogrammetric mapping \cite{bib:Long2015}, and others.

Usually, BPDN solvers are either first- or second-order methods. First-order methods are typically based on linear approximations. Examples include iterative shrinkage thresholding methods (ISTA), alternating direction method of multipliers (ADMM), and coordinate descent (CD). As for second-order methods, they are often computationally expensive. An example of a second-order method is the primal-dual interior-point method (PDIPM), which has computationally expensive iterations. In any wise, sparse reconstruction based methods are computationally much more expensive than classic linear methods. Yet, specific structures of the sensing matrix $\mathbf{A}$ can be exploited for faster solutions.

In this work, we address tomographic SAR (TomoSAR), for which $\mathbf{A}$ is an irregular Fourier transform matrix with a typical matrix size of ca. 100 times 1 million. For instance, $\mathbf{A}$ has the dimension $n \times m$, where $n$ is equal to the number of interferograms in the application of TomoSAR and $m$ is equal to the amount of discretization. The typical value of $n$ is from 20 to 100, and $m$ is above 1 million. When multiplied, they indicate the amount of discretization along each dimension, such as elevation, seasonal motion and linear deformation. Besides TomoSAR, our findings and algorithms are applicable to further examples, such as SAR focusing, inverse SAR, and underground sonar imaging for DOA estimation.  In TomoSAR, it is demonstrated that basis pursuit denoising approach based on $L_1$ regularized least squares, such as ``scale-down by $L_1$ norm minimization, model selection, and estimation reconstruction" (SL1MMER algorithms, pronounced ``slimmer") proposed in \cite{bib:Zhu2010a}, can achieve significant super-resolution \cite{bib:Zhu2012b}\cite{bib:Zhu2015}, compared to classic linear methods \cite{bib:Fornaro2005}\cite{bib:Zhu2010b}. Yet the downside of this method is its computational cost. For example, to reconstruct one scene covered by a radar satellite image, like in TerraSAR-X high-resolution spotlight mode, about 20 million problems of the abovementioned size should be solved, which makes it infeasible for large-scale processing. In \cite{bib:Wang2014}, Wang et al. proposed an efficient approach to address this issue, which uses the well-established and computationally efficient persistent scatterer interferometry \cite{bib:Ferretti2001} to obtain a prior knowledge of the estimates, followed by the linear method and L1LS-based SL1MMER algorithm applied to pre-classified different groups of pixels. This approach speeds up the processing, but only to the extent of reducing the percentage of pixels that requires sparse reconstruction. This is to say, if $10\%$ of the pixels will be processed by SL1MMER (i.e., solving L1LS), the whole processing can only be sped up by up to a factor of ten. In other words, the strategy is to only use theses algorithms for pixels where super-resolution is needed. For the rest of the pixels, processing will be done with fast algorithms, e.g., linear estimators. Since the computational time of linear estimators is almost negligible compared to sparse reconstruction algorithms, in the end, the percentage of pixels demanding super-resolving is decisive to the possible extent of speeding up. In this work, we want to speed up the sparse reconstruction algorithms, which is currently causing a bottleneck.

The main contributions of this paper are listed as follows;
\begin{itemize}
\item A novel approach “randomized blockwise proximal gradient” (RBPG) has been proposed to solve the complex-valued sparse optimization problem in radar remote sensing.
\item Systematic performance evaluation of the proposed approach has been carried out using both simulated data and real data with the application of tomographic SAR. The results show that it can maintain the accuracy and super-resolution power of second order sparse reconstruction methods and dramatically speed up the whole processing by one or two orders.
\item Operational-level processing for a large-scale problem has been carried out, which is demonstrated by an accurate 4-D point cloud reconstruction over a very large area -- the whole city of Munich.
\end{itemize}

The paper is organized as follows: In section II, the high-dimensional SAR imaging model and TomoSAR inversion are introduced; In section III, the SL1MMER algorithms are reviewed, and a novel approach for its sparse optimization procedure is introduced; The experiments, using simulated data and real data, are presented in section IV; Finally, conclusions are given in section V.

\section{SAR Imaging}
In this section, we first introduce the high-dimensional SAR imaging model for TomoSAR. Furthermore, we compare different TomoSAR inversion approaches.
As an extension of TomoSAR, differential synthetic aperture radar tomography (D-TomoSAR) uses multi-baseline, multi-temporal SAR acquisitions for reconstructing the 3D distribution of scatterers and their motion \cite{bib:Lombardini2003}\cite{bib:Fornaro2009}\cite{bib:Zhu2012a}. The D-TomoSAR system model can be expressed as follows:

\begin{figure}
\centering
\includegraphics[width=0.5\textwidth]{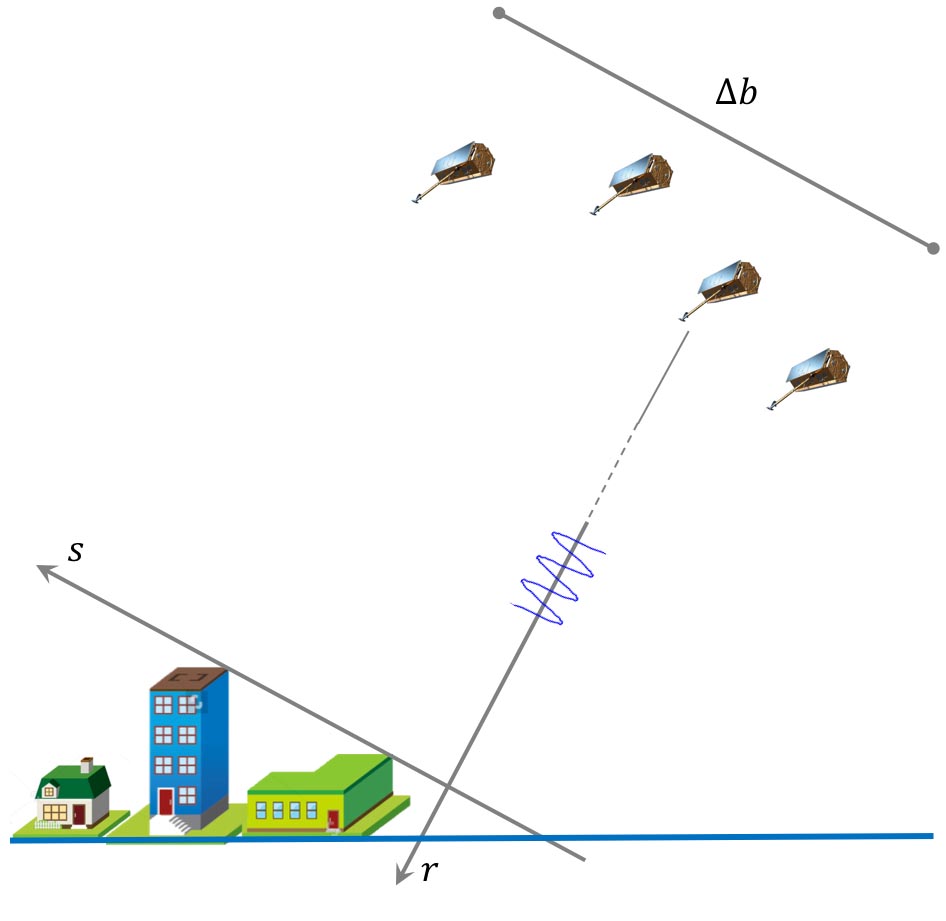}
\caption{TomoSAR imaging geometry with an artistic view of TerraSAR-X/TanDEM-X}
\label{fig:tomosar_geometey}
\end{figure}

\begin{equation}
g_n = \int_{\Delta s}\gamma(s)\exp(j2\pi(\xi_ns+2d(s,t_n)/\lambda))ds
\end{equation}

where $g_n$ is the complex-valued measurement at an azimuth-range pixel for the $n$th acquisition at time $t_n (n = 1,2,...,N)$. $\gamma(s)$ represents the reflectivity function along elevation $s$ with an extent of $\Delta s$, $\xi_n = 2b_n/(\lambda r)$ is the spatial frequency proportional to the respective aperture position (baseline) $b_n$, $\lambda$ is the wavelength and $r$ is the range. $d(s,t_n)$ is the line-of-sight (LOS) motion as a function of elevation and time. The motion relative to the master acquisition may be modeled using a linear combination of the $M$ base function $\tau_m(t_n)$

\begin{equation}
d(s,t_n) = \sum_{m=1}^M p_m(s)\tau_m(t_n)
\end{equation}

where $p_m(s)$ is the corresponding motion coefficient to be estimated and $\tau_m(t_n)$ are the temporal frequencies. The choice of the base functions depends on the underlying physical motion process. Therefore, we generalize it in the multicomponent model:

\begin{equation}
\begin{aligned}
g_n &= \int...\int\int\gamma(s)\delta(p_1-p_1(s),...,p_M-p_M(s)) \\
&\exp(j2\pi(\xi_ns+\eta_{1,n}p_1+...+\eta_{M,n}p_M))dsdp_1...dp_M
\end{aligned}
\end{equation}

The inversion of the system model provides retrieval of the elevation and deformation information, even of multiple scatterers inside an azimuth-range resolution cell, thus obtaining a high-dimensional map of scatterers.
In the presence of noise $\bf{\varepsilon}$, the discrete-TomoSAR system model can be rewritten

\begin{equation}
\bf{g}=\bf{R}\bf{\gamma}+\bf{\varepsilon}
\label{equ:tomosar_basic}
\end{equation}

where $\bf{g}$ is the measurement vector with $N$ elements, and $\bf{\gamma}$ is the reflectivity function along elevation uniformly sampled at $s_l (l=1,2,...,L)$. $\bf{R}$ is an $N \times L$ irregularly sampled discrete Fourier transformation mapping matrix.

Theoretically speaking, we would ideally solve Eq. (\ref{equ:tomosar_basic}) by $L_0$ minimization, which would give the correct solution, but, unfortunately, the $L_0$ minimization problem is NP-hard. For $L \gg N$ (i.e., with $\bf{\gamma}$ sufficiently sparse), it can be shown that the $L_1$ norm minimization leads to nearly the same result as the $L_0$ minimization.

\section{Methodology}
\subsection{Review of SL1MMER}
In \cite{bib:Zhu2010a}, Zhu et al. proposed the SL1MMER algorithm. They demonstrated its super-resolution power and robustness for spaceborne tomographic SAR in \cite{bib:Zhu2012c}\cite{bib:Zhu2014b}. The SL1MMER algorithm improves the CS algorithm and estimates these parameters in a very accurate and robust way. It consists of three main steps: (1) an L1LS minimization, (2) model selection, and (3) parameter estimation. Among all the steps, L1LS minimization is the most time-consuming one. In case there is no prior knowledge about the number of scatters, and in the presence of measurement noise, it can be approximated by

\begin{equation}
\hat{\bf{\gamma}} = \arg \min_{\bf{\gamma}} \{ \Vert \bf{R}\bf{\gamma} - \bf{g} \Vert^2_2 + \lambda \Vert \bf{\gamma} \Vert_1 \}
\label{equ:opt_l1lsp}
\end{equation}

Generic methods for non-differentiable convex problems, such as the ellipsoid method or subgradient methods \cite{bib:Shor1985}\cite{bib:Polyak1987}, can be used to solve Eq. (\ref{equ:opt_l1lsp}). These methods are often very slow. The equation (\ref{equ:opt_l1lsp}) can be transformed to a convex quadratic problem, with linear inequality constraints. The equivalent quadratic program (QP) can be solved by standard convex optimization methods such as interior-point methods. However, the data of InSAR is complex-valued, which requires the use of the second order cone program (SOCP), instead of QP, for solving Eq. (\ref{equ:opt_l1lsp}). In \cite{bib:Zhu2010a}, the second order method PDIPM with self-dual embedding techniques was adopted to solve the SOCP. This is computationally expensive and is difficult to extend to large scales. To make TomoSAR processing fit for high throughput or operational use, a fast L1LS solver is crucial.

\subsection{Randomized Blockwise Proximal Gradient Algorithms}
In this section, we propose a novel approach for solving L1LS minimization, which can retain the super-resolution power of the standard BPDN solver and extremely speed up the processing for matrix \textbf{A} of the random Fourier transform as used in TomoSAR.

Our unconstrained optimization problems with an objective function can be split into the convex differentiable part and convex non-differentiable part, leading to the so-called proximal gradient (PG) method. The PG method is used for optimization of an unconstrained problem with an objective function $F({\bf{x}})$ split in two components. We consider the following problem,

\begin{equation}
\min_{\bf{x}} F({\bf{x}}) =  f({\bf{x}}) + r({\bf{x}})
\label{equ:opt_base}
\end{equation}

where $f(\bf{x})$ is the convex differentiable function, and $r(\bf{x})$ is the convex and non-differentiable regularization function. The iterative approach to solve (\ref{equ:opt_base}) can be written as

\begin{equation}
{{\bf{x}}^{k+1}}= \arg \min \Big( \langle \nabla f({\bf{x}}^k), {\bf{x}}-{\bf{x}}^k \rangle + \frac{1}{2\alpha_k}\Vert {\bf{x}}-{\bf{x}}^k \Vert_2^2 + r(\bf{x}) \Big)
\label{equ:opt_sol}
\end{equation}

where $\nabla f$ is the partial gradient of function $f$. The proximal gradient formulation is

\begin{equation}
{\bf{x}}^{k+1} ={\rm prox}_{\alpha_kr}({\bf{x}}^k - \alpha_k\nabla f({\bf{x}}^k))
\end{equation}

where $\alpha_k > 0$ is the step size, which can be constant or determined by line search. For $r({\bf{x}}) = \Vert {\bf{x}} \Vert_1$, the proximal operator can be chosen as soft-thresholding:

\begin{equation}
{\rm prox}_{\alpha_kr}(x) =
\begin{cases}
x - \alpha_k, & x > \alpha_k \\
0, & -\alpha_k \leq x \leq \alpha_k \\
x + \alpha_k, & x < \alpha_k
\end{cases}
\end{equation}

Proximal gradient algorithms can be accelerated by using Nesterov's method \cite{bib:Nesterov1983} in the following way:

\begin{align}
{\bf{y}}^{k+1} &= {\bf{x}}^k + \theta_k(\dfrac{1}{\theta_{k-1}}-1)({\bf{x}}^k - {\bf{x}}^{k-1}) \\
{\bf{x}}^{k+1} &= {\rm prox}_{\alpha_kr}({\bf{y}}^{k+1} - \alpha_k\nabla f({\bf{y}}^{k+1}))
\end{align}

where $\theta_k$ is chosen as $2/(k+1)$. The convergence rate of the basic PG algorithms is improved to $O(1/k^2)$ by the extrapolation. In order to further accelerate the algorithms, a randomized block coordinate is adopted. As shown in \cite{bib:Peng2016}\cite{bib:Xu2015}, by applying block coordinate techniques, the equation (\ref{equ:opt_sol}) can be written as

\begin{eqnarray}
{\bf{x}}_{i_k}^{k+1} = &\arg \min \Big(\langle \nabla f_{i_k}({\bf{x}}_{i_k}^k), {\bf{x}}_{i_k}-{\bf{x}}_{i_k}^k \rangle \nonumber \\
&+ \frac{1}{2\alpha_{i_k}^k}\Vert {\bf{x}}_{i_k}-{\bf{x}}_{i_k}^k \Vert_2^2 + r_{i_k}(\bf{x}) \Big)
\end{eqnarray}

where $i_k$ is the index of the block. The choice of the update index $i_k$ for each iteration is crucial for good performance. Often, it is easy to switch index orders. However, the choice of index affects convergence, possibly resulting in faster convergence or divergence. In this work, we chose the randomized variants scheme, which has strengths, such as less memory consumption, good convergence performance and empirical avoidance of the local optimal. $i_k$ is chosen randomly following the probability distribution given by the vector

\begin{equation}
P_{i_k} = \dfrac{L_{i_k}}{\sum_{j=1}^J L_j}, \quad i_k = 1,...,J
\end{equation}

where $L_{i_k}$ is the Lipschitz constant of $\nabla_{i_k}f({\bf x})$, the gradient of $f({\bf x})$ with respect to the $i_k$-th group (in our case, $L = ||A^TA||$). However, setting $\alpha_k = 1/L$ usually results in very small step sizes. Consequently, the time step $\alpha_k$ is adaptively chosen by using backtracking line search method in \cite{bib:Xu2017}\cite{bib:Peng2013}.

The step length $\alpha_k$ is determined iteratively by multiplication of a factor, $C_{\alpha} \in (0,1)$, until the following holds:

\begin{equation}
f(x') \le f(x) + \nabla f(x)^T (x'-x) + \dfrac{1}{2\alpha_k}||x'-x||^2
\label{equ:line_search}
\end{equation}

This condition ensures that the value $f(x')$ of $f$ at the new point $x'$ is smaller than the value of quadratic approximation at the point $x$. The framework of our method is given in Algorithm \ref{alg:rbpg}.

\begin{algorithm}
 \caption{RBPG with backtracking}
 \label{alg:rbpg}
 \begin{algorithmic}[1]
 \renewcommand{\algorithmicrequire}{\textbf{Init:}}
 \REQUIRE $x^{(0)}, y^{(0)}=0$; for $k \ge 1$, repeat the steps
  \FOR {$k = 1,2,...,N_K$}
  \STATE
  \STATE $i_k \leftarrow  P_{i_k} = \frac{L_{i_k}}{\sum_{j=1}^J L_j}$
  \STATE ${\bf{y}}_{i_k}^{{}k+1} \leftarrow {\bf{x}}_{i_k}^k + \theta_k(\frac{1}{\theta_{k-1}}-1)({\bf{x}}_{i_k}^k - {\bf{x}}_{i_k}^{k-1})$
  \STATE $\bar{\bf{x}}_{i_k}^{k+1} \leftarrow {\rm prox}_{\alpha_kr}({\bf{y}}_{i_k}^{k+1} - \alpha_k\nabla f({\bf{y}}_{i_k}^{k+1}))$
  \STATE
  \WHILE {(Eq. (\ref{equ:line_search}) is fulfilled)}
  \STATE $\alpha_k = C_{\alpha} \cdot \alpha_k$
  \STATE repeat steps $4,5$
  \ENDWHILE
  \STATE
  \IF {($F(\bar{\bf{x}}_{i_k}^{k+1}) \le F({\bf{x}}_{i_k}^k)$)}
  \STATE ${\bf{x}}_{i_k}^{k+1} = \bar{\bf{x}}_{i_k}^{k+1}$
  \ELSE
  \STATE ${\bf{x}}_{i_k}^{k+1} = {\bf{x}}_{i_k}^k$
  \ENDIF
  \STATE
  \ENDFOR
 \end{algorithmic}
 \end{algorithm}

 \begin{figure*}
 \centering
 \subfloat[]{\includegraphics[width=0.4\textwidth]{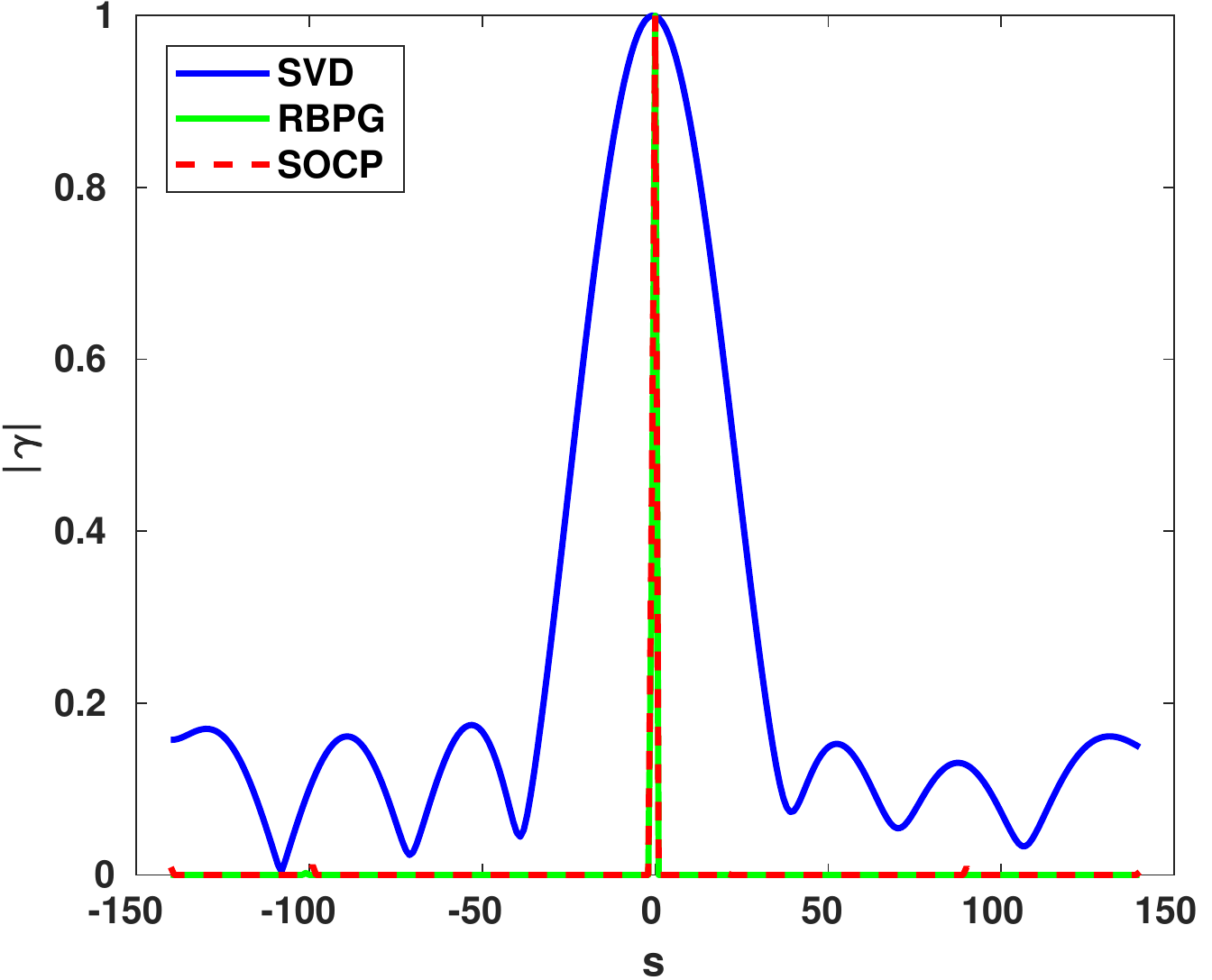}}
 \qquad
 \subfloat[]{\includegraphics[width=0.4\textwidth]{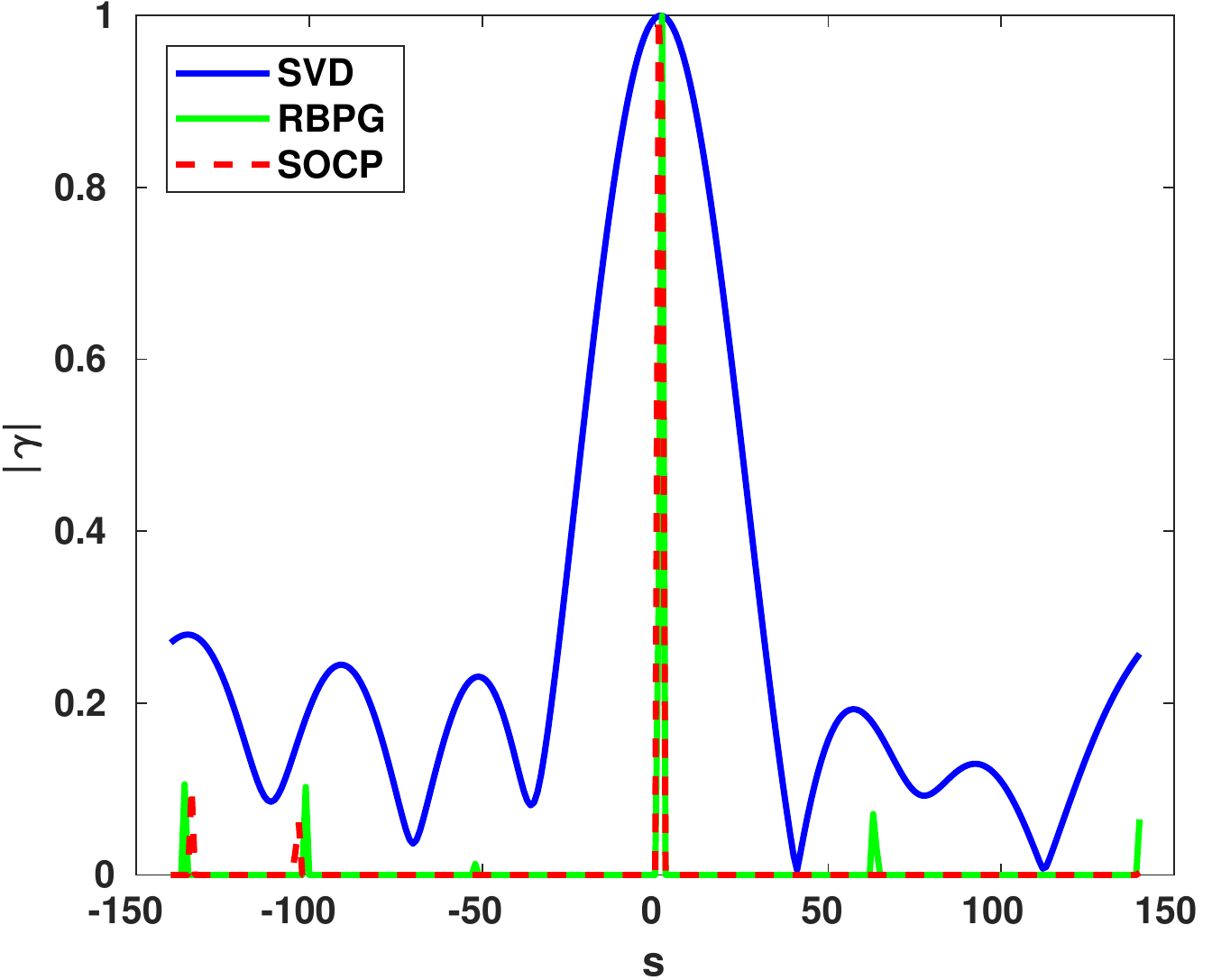}}
 \caption{Performance comparison between SVD (blue), RBPG (green) and SOCP (red) on simulated data with single scatterer. (a) SNR = 10 dB (b) SNR = 3 dB}
 \label{fig:sim_svd_rbgp_socp_1}
 \end{figure*}

\subsection{Complexity Analysis}
The complexity of each algorithms is analyzed in this section. $O(1)$ is assumed as the computational complexity for one multiplication.

Among all approaches, SVD-Wiener (Eq.(11) in \cite{bib:Zhu2010b}) needs the least complexity $O(N^2L_s)$, which pays with the price of non super-resolution. $N$ is the number of acquisitions. $L_s$ is the multiplication of discretization levels in three dimensions, namely elevation direction, linear motion direction and seasonal motion direction. The typical range for each direction can be 200-400, 20-40, or 30-40. In SL1MMER, each iteration of the PDIPM is dominated by the cost of computing the PD search direction from the Newton system. This leads to the computational complexity $O(\epsilon_pK_pL^3)$, where $L$ is the multiplication of discretization levels in three dimensions for sparse reconstruction, which is about 2-10 times smaller than $L_s$. $K_p$ is the number of iterations of PDIPM (approximately 10-20), and $\epsilon_p$ is the acceleration factor of PDIPM due to different techniques, such as the preconditioned conjugate gradient (0.1-1.0). The main computational cost of RBPG is due to Eq. (13), which requires at least $O(K_rM_rL^2)$, where $K_r$ is the number of iterations of RBPG (approximately 50-100) and $M_r$ is the number of multiplications of a specific matrix in each iteration. According to the computational complexity, RBPG should be ten to several hundred times faster than SOCP.

\section{Experiments}
\subsection{Simulation}
In this section, we compare the RBPG approach to the SOCP and SVD approaches using simulated data. The inherent (Rayleigh) elevation resolution $\rho_s$ of the tomographic arrangement is related to the elevation aperture extent $\Delta b$ \cite{bib:Zhu2012b}

\begin{equation}
\rho_s = \dfrac{\lambda r}{\Delta b}
\end{equation}
The normalized distance is defined as
\begin{equation}
\kappa = \dfrac{s}{\rho_s}
\end{equation}

For the first test case, only one scatterer is placed at $s = 0$, and two SNR are chosen: 10 dB and 3 dB. Fig. \ref{fig:sim_svd_rbgp_socp_1} shows a performance comparison between SVD, RBPG and SOCP on simulated data with a single scatterer. As one can see, all methods can detect the position of the single scatterer, although the reflectivity profile reconstructed by SVD has more sidelobe than the others.

\begin{figure*}
\centering
\subfloat[]{\includegraphics[width=0.3\textwidth]{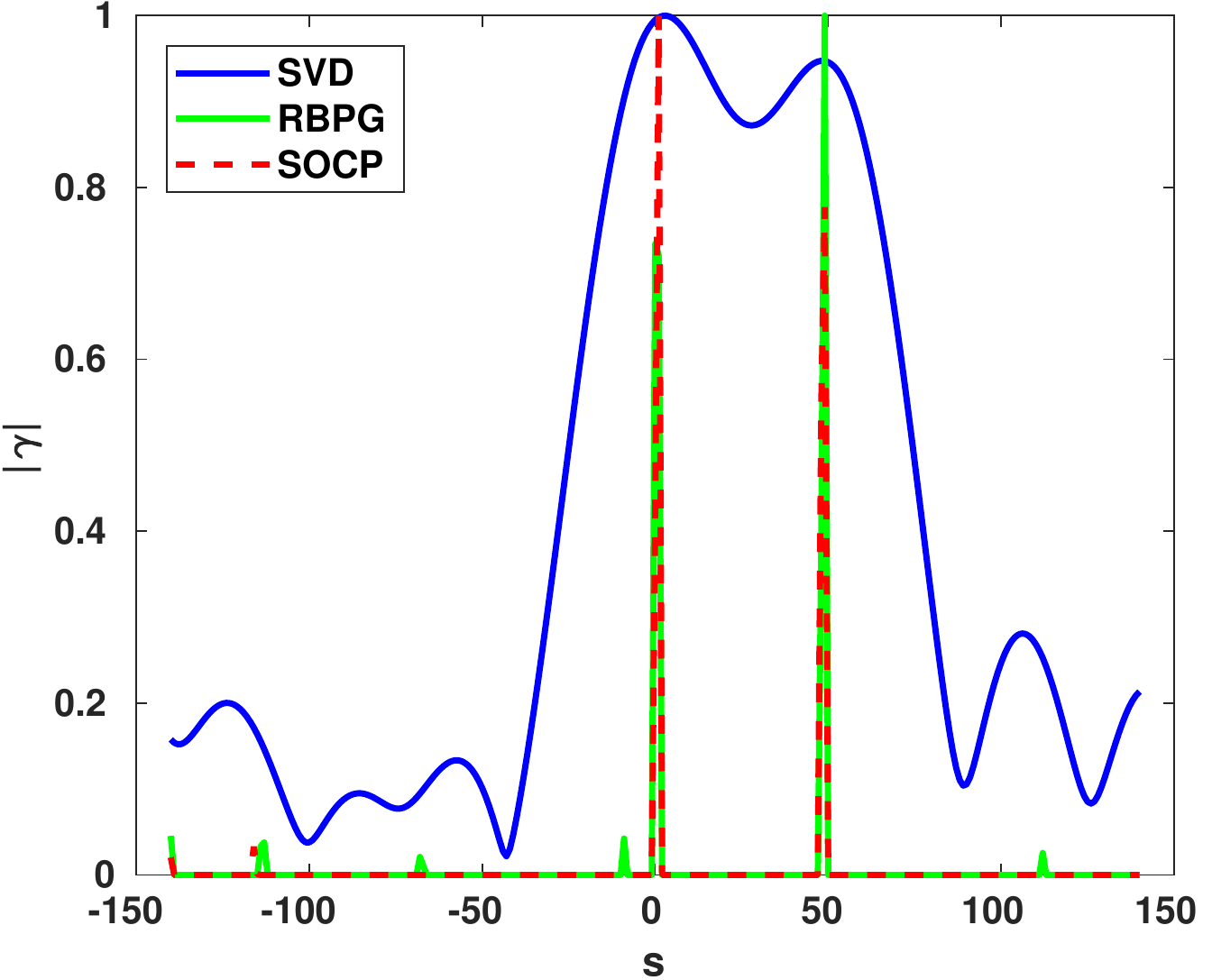}}
\subfloat[]{\includegraphics[width=0.3\textwidth]{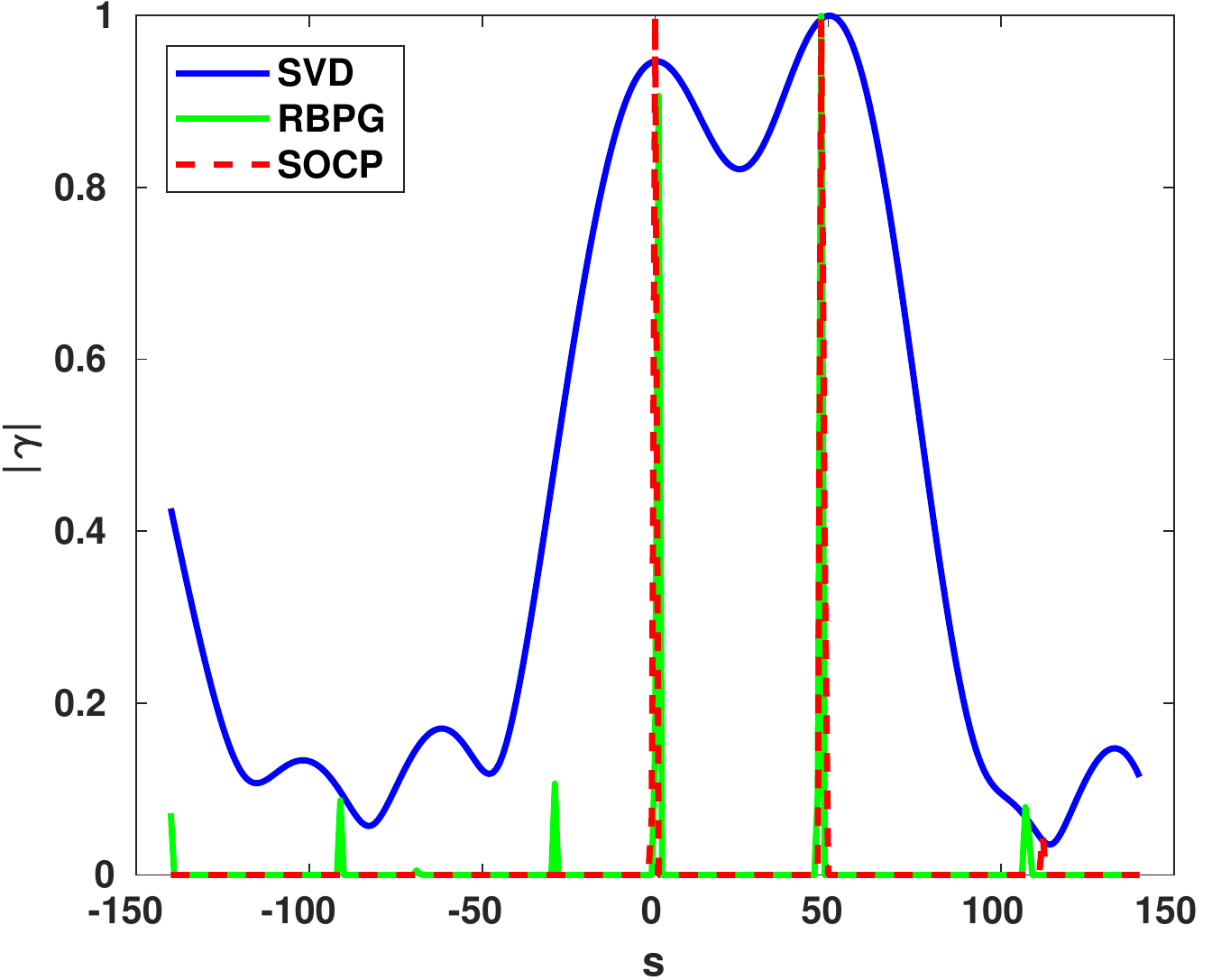}}
\subfloat[]{\includegraphics[width=0.3\textwidth]{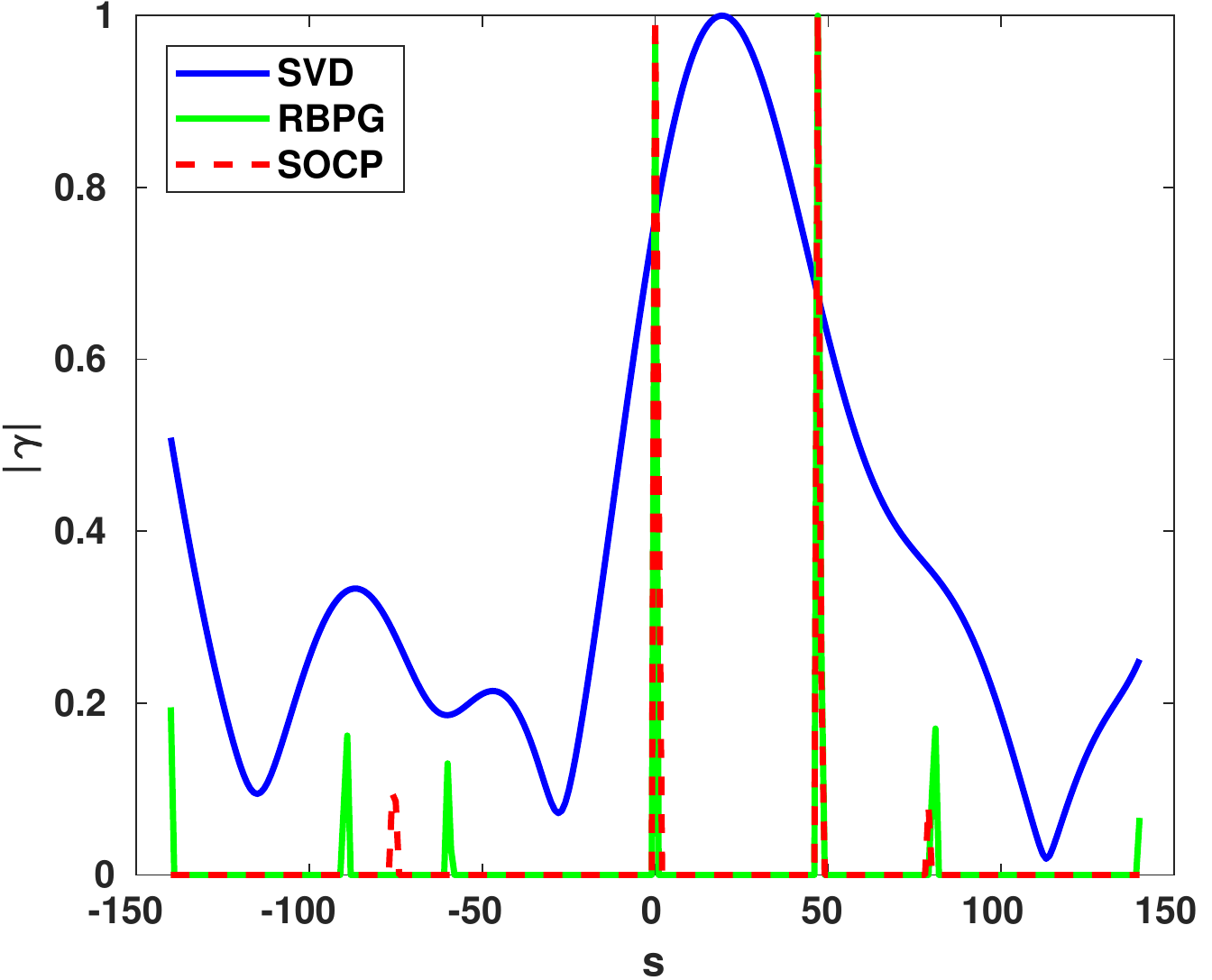}}
\hfil
\subfloat[]{\includegraphics[width=0.3\textwidth]{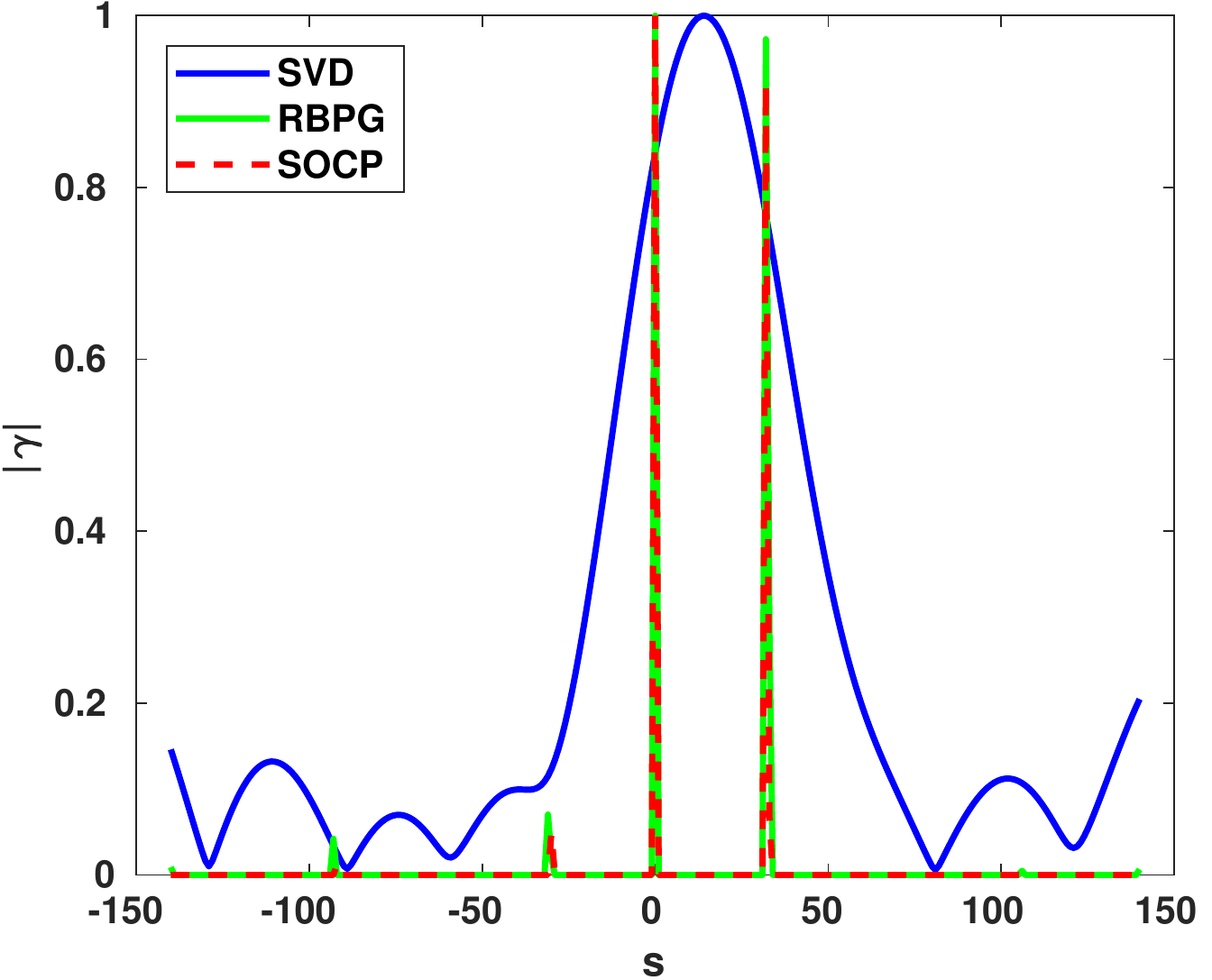}}
\subfloat[]{\includegraphics[width=0.3\textwidth]{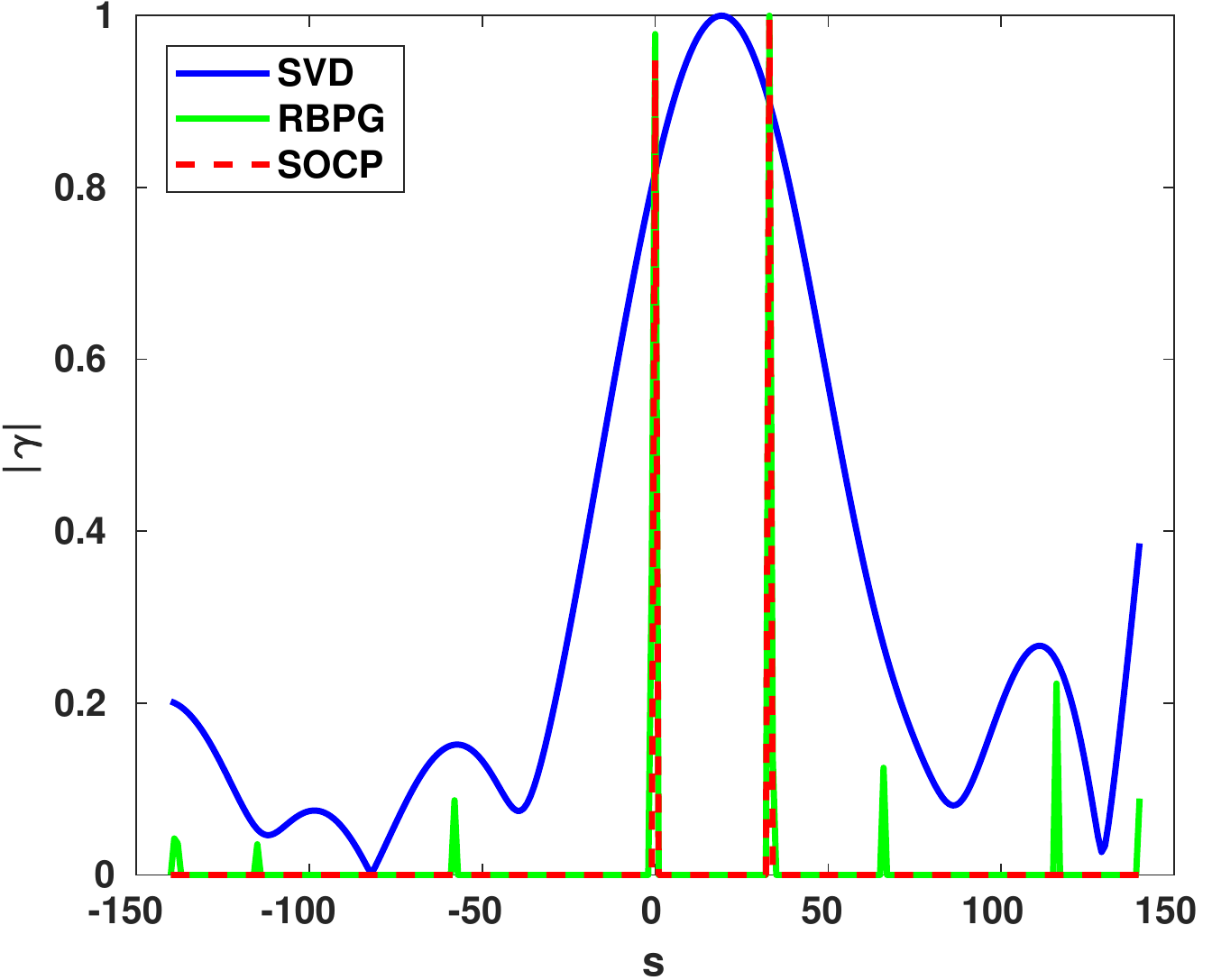}}
\subfloat[]{\includegraphics[width=0.3\textwidth]{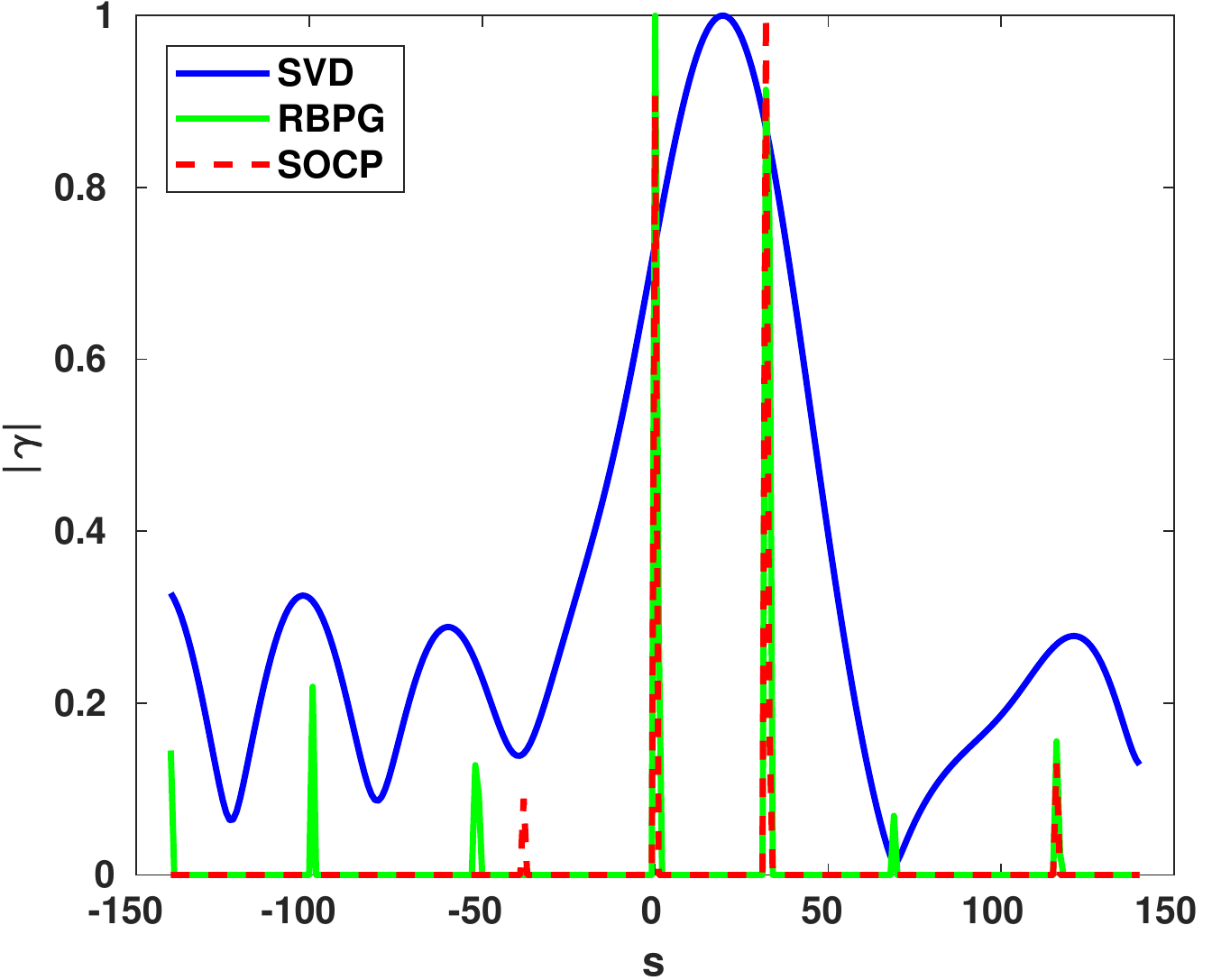}}
\hfil
\subfloat[]{\includegraphics[width=0.3\textwidth]{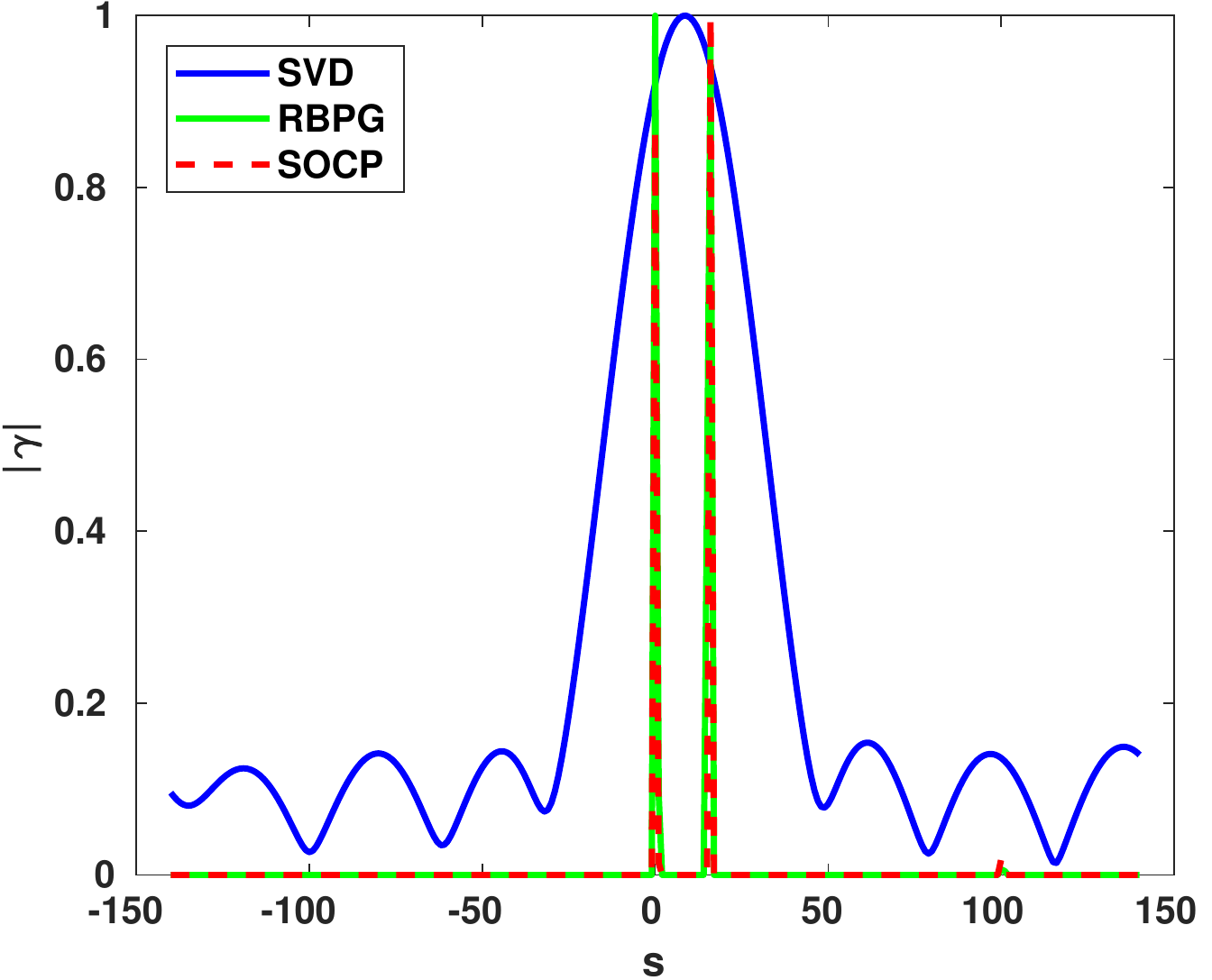}}
\subfloat[]{\includegraphics[width=0.3\textwidth]{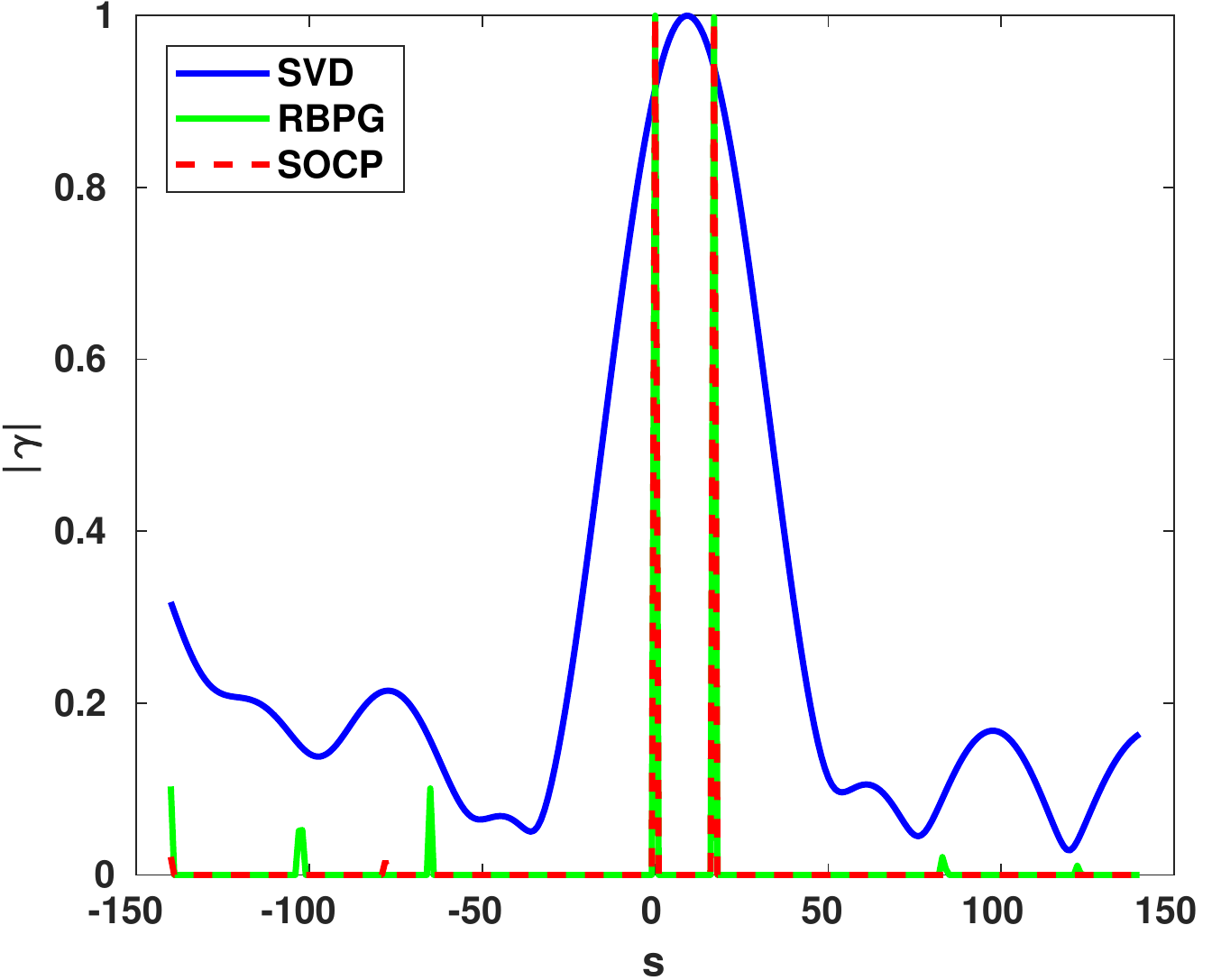}}
\subfloat[]{\includegraphics[width=0.3\textwidth]{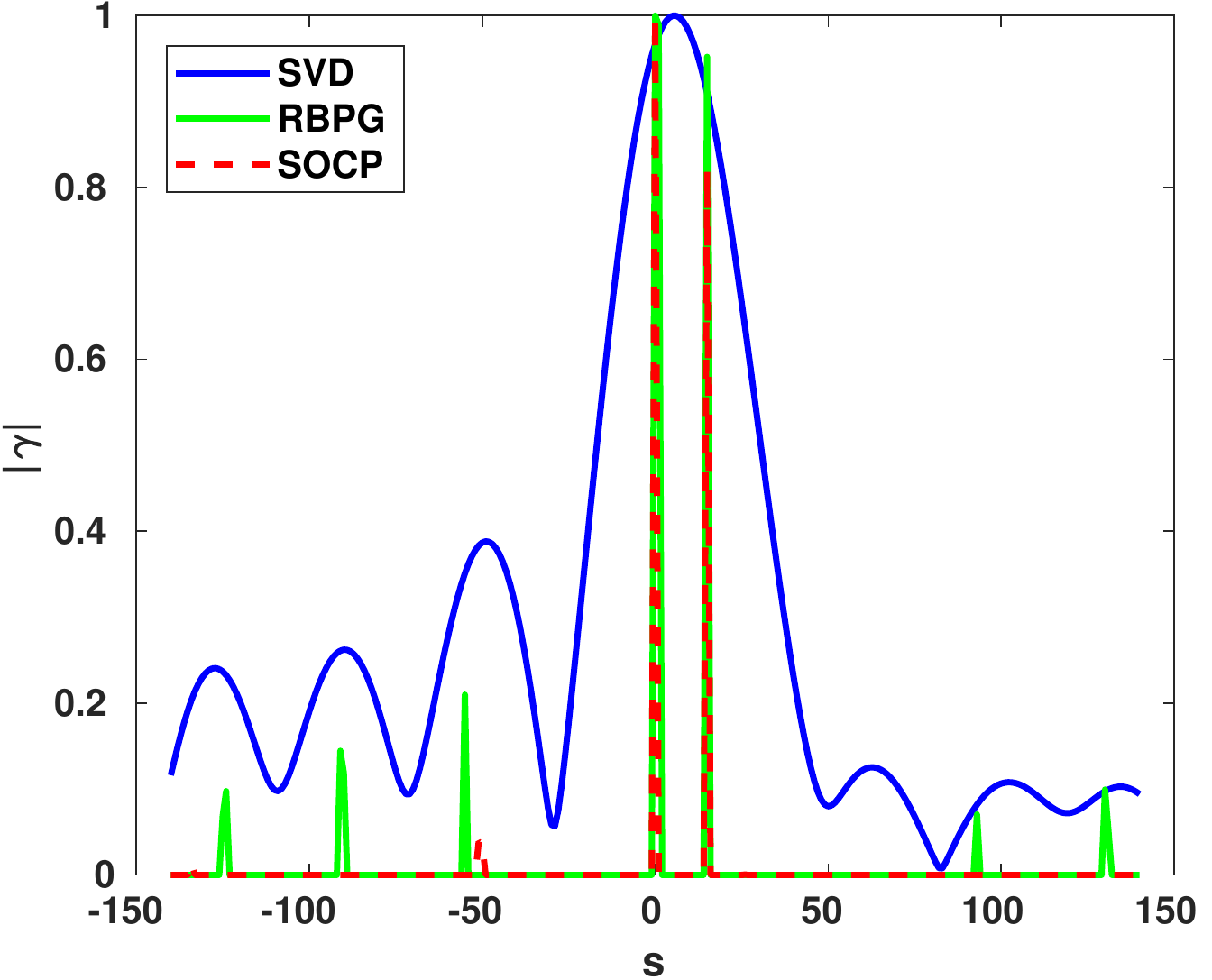}}
\hfil
\subfloat[]{\includegraphics[width=0.3\textwidth]{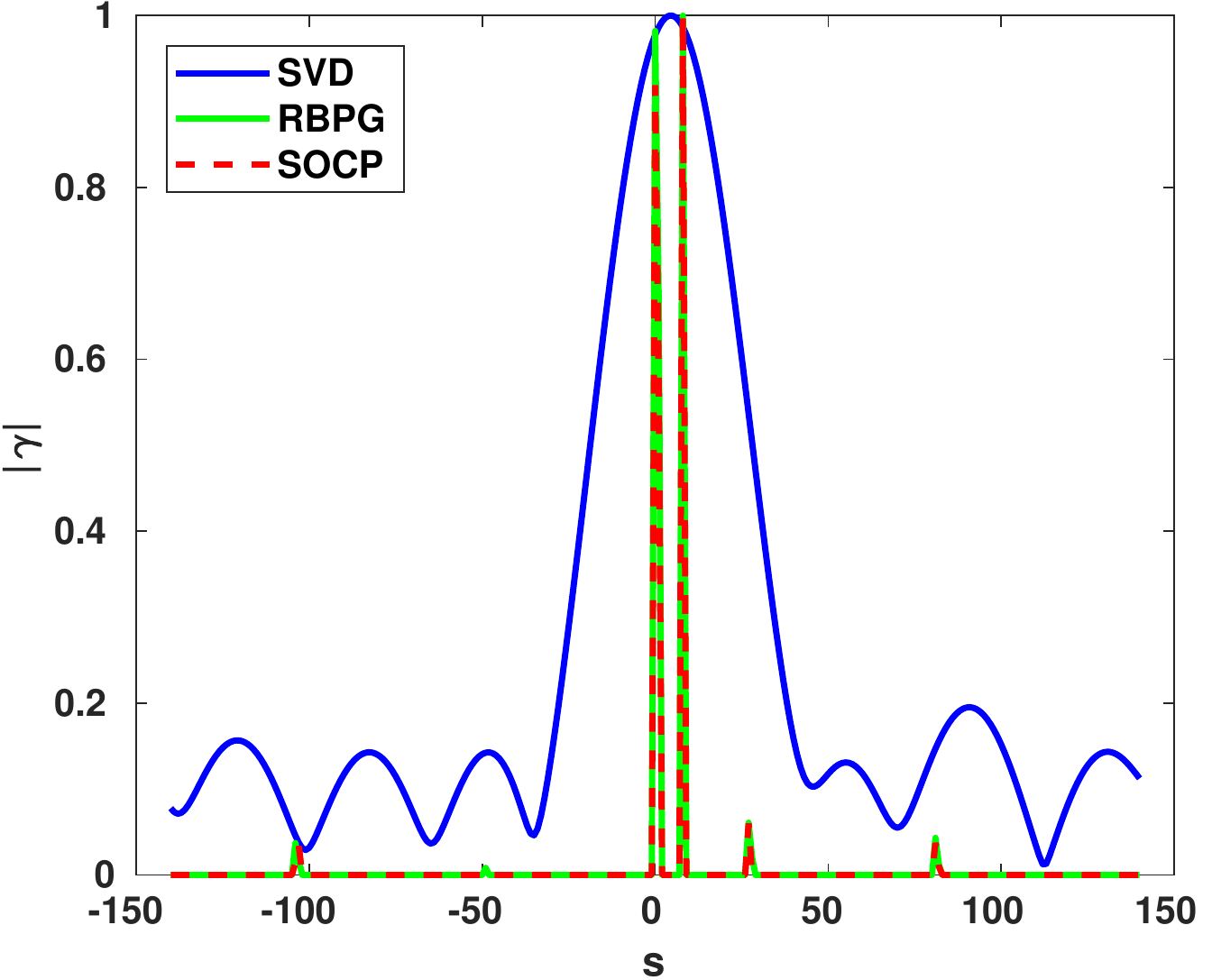}}
\subfloat[]{\includegraphics[width=0.3\textwidth]{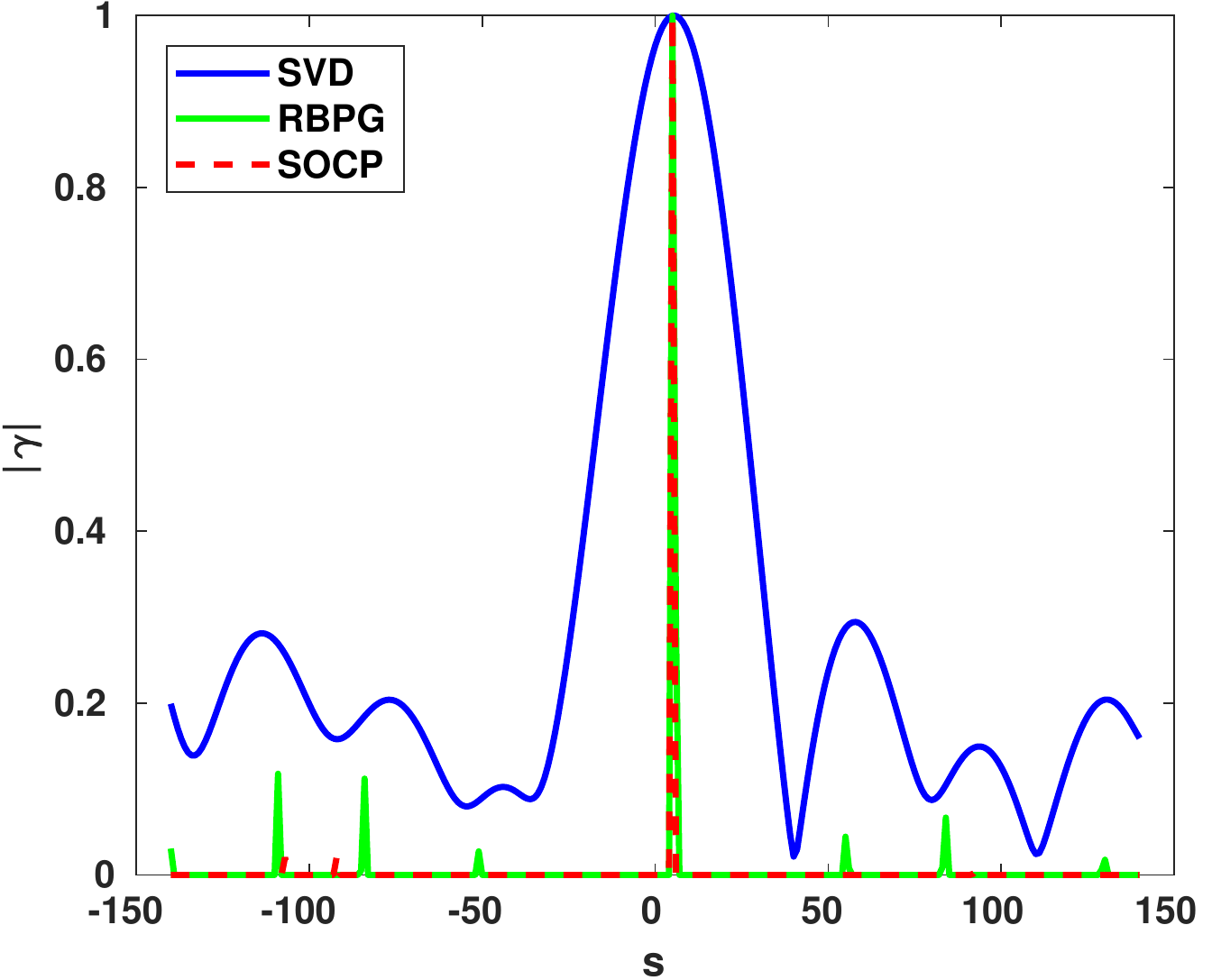}}
\subfloat[]{\includegraphics[width=0.3\textwidth]{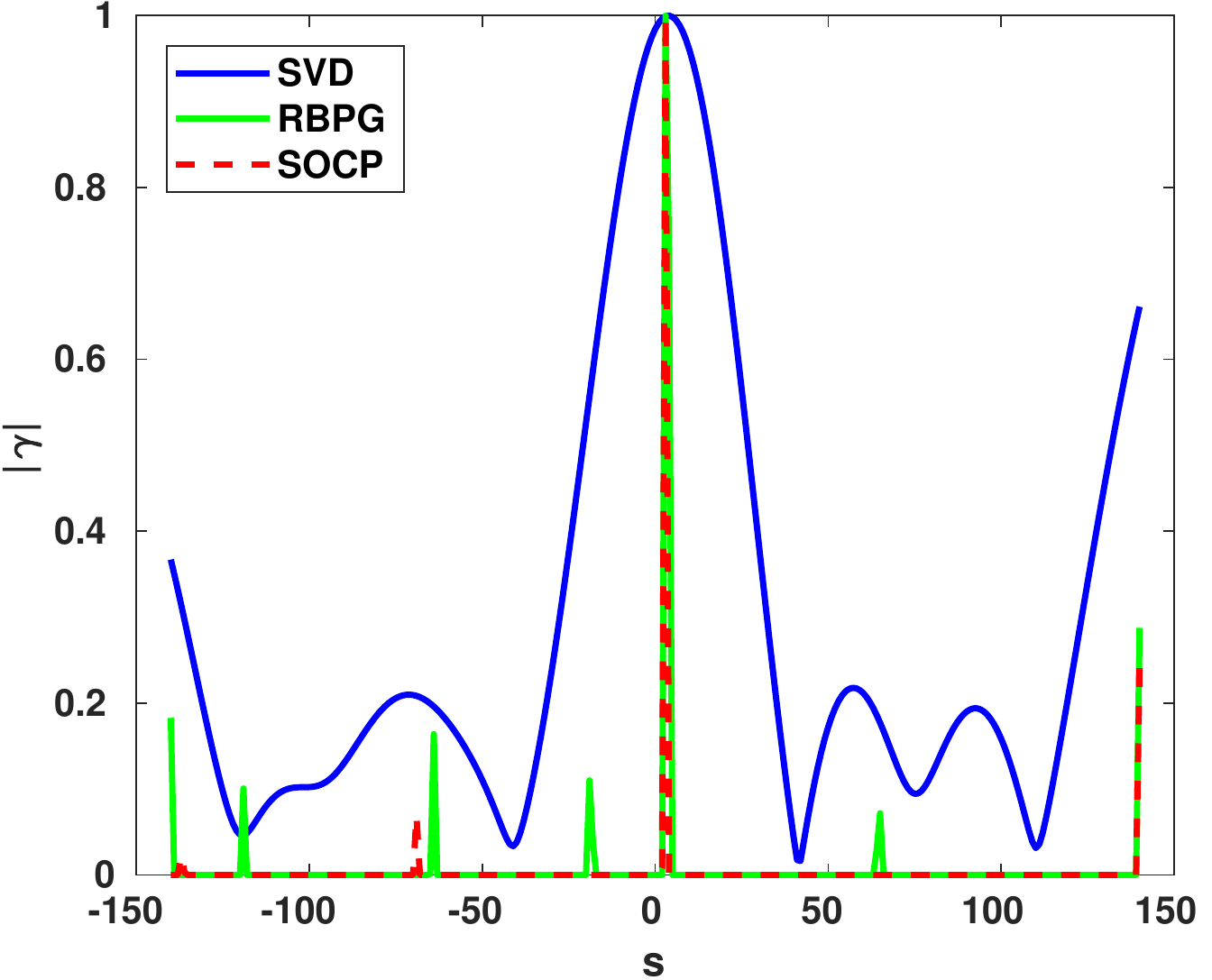}}
\caption{Performance comparison between SVD (blue), RBGP (green) and SOCP (red) on simulated data with two scatterers. (a)(d)(g)(j) with SNR = 10 dB, (b)(e)(h)(k) with SNR = 3 dB, (c)(f)(i)(l) with SNR = 0 dB, and the normalized distance $\kappa = 1.2, 0.8, 0.4, 0.2$ from top to bottom.}
\label{fig:sim_svd_rbgp_socp}
\end{figure*}

For the double scatterers' case, we assume the situation with two scatterers inside an azimuth-range pixel: one scatterer located at the building facade and another from the ground with four different normalized distances : $\kappa = 1.2,0.8,0.4,0.2$; and a number of acquisitions, $N = 29$. Fig. \ref{fig:sim_svd_rbgp_socp} shows the comparison of the reconstructed reflectivity profiles along the elevation direction by SVD (blue solid lines), RBPG (green dash lines) and SOCP (red solid lines), where the x-axis is the absolute value of normalized reflectivity $\gamma$, and the y-axis is the elevation $s$.

\begin{figure*}
\centering
\subfloat[]{\includegraphics[width=0.45\textwidth]{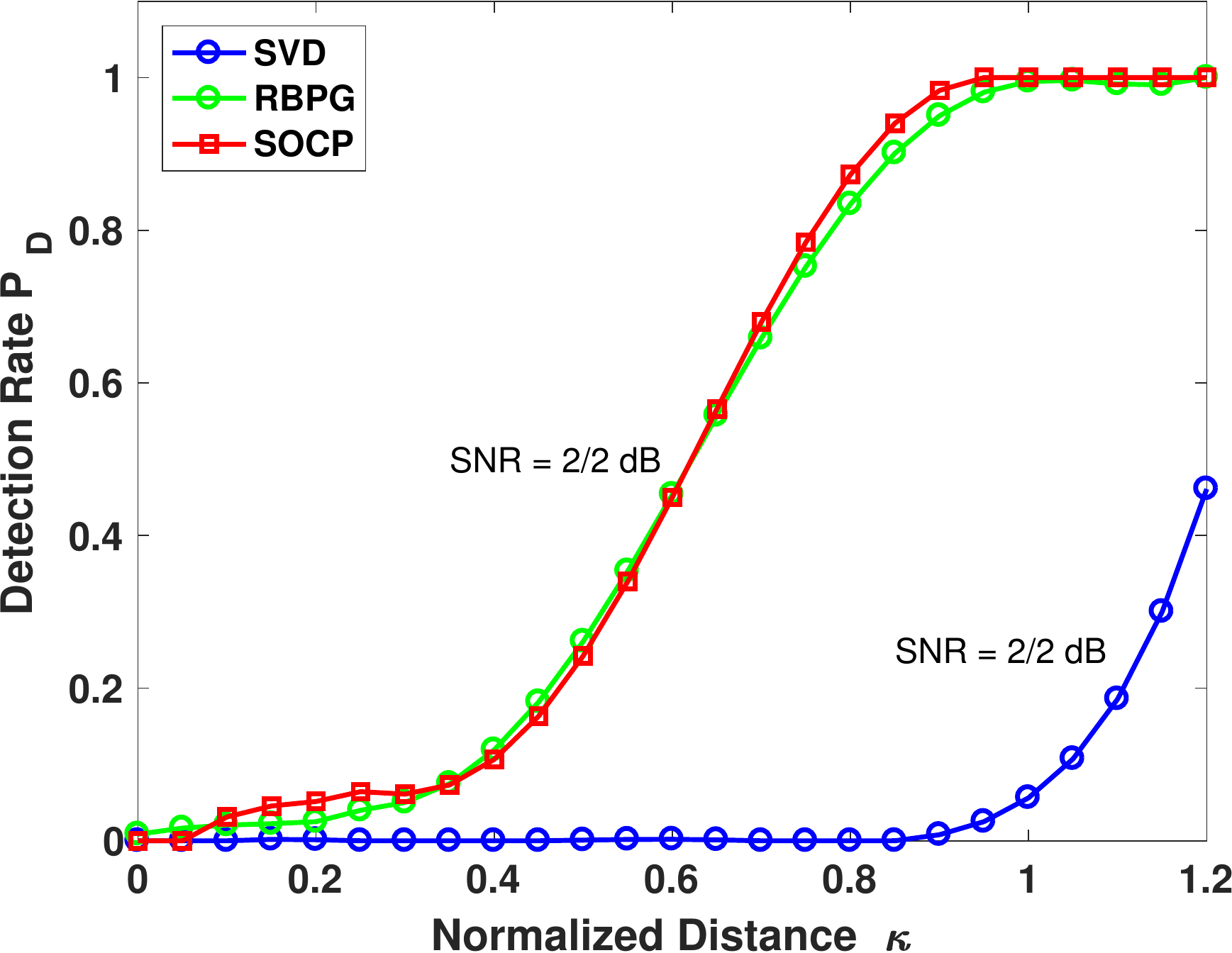}}
\qquad
\subfloat[]{\includegraphics[width=0.45\textwidth]{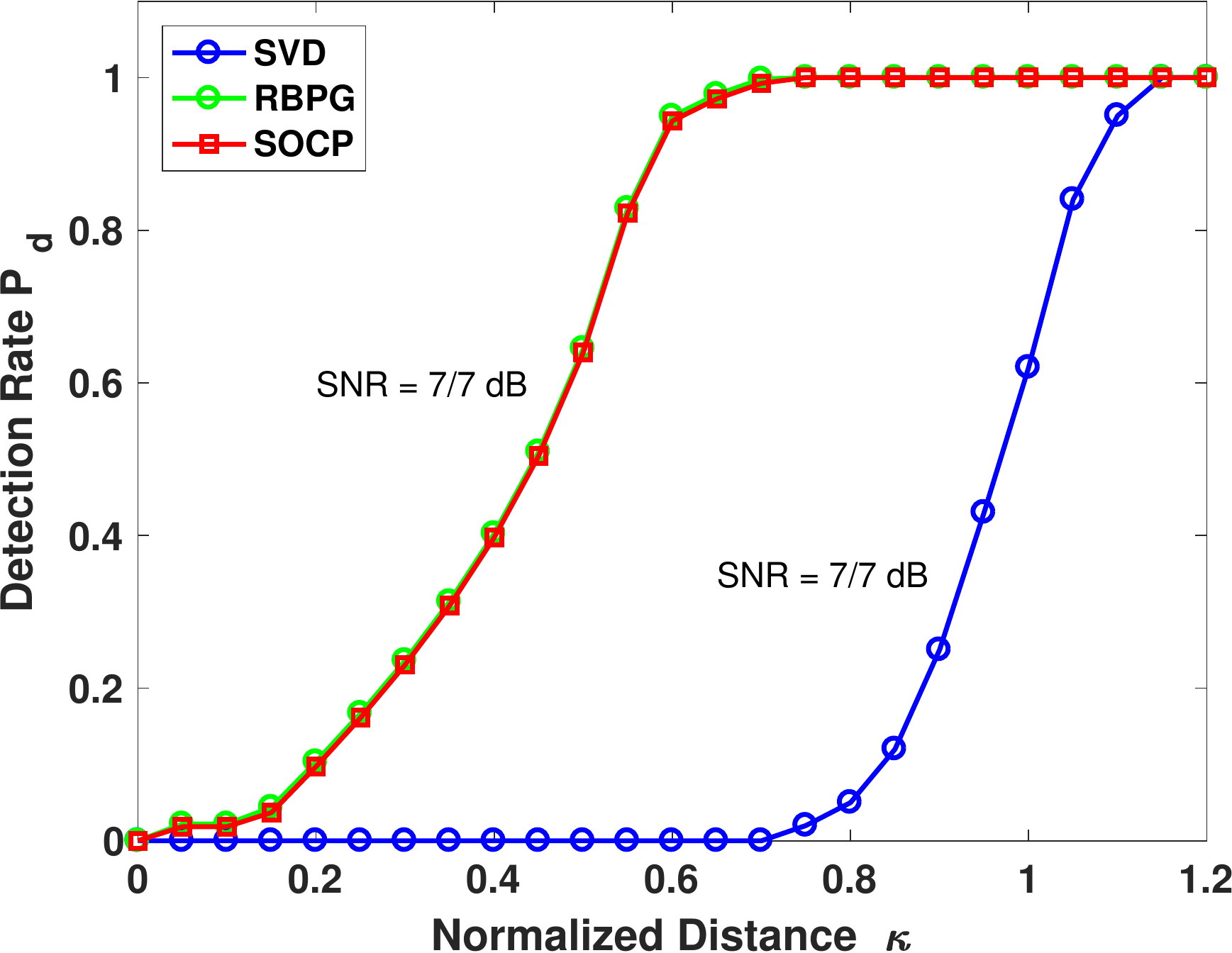}}
\caption{Detection rate as a function of normalized distance $\kappa$ using SVD (blue), RBPG (green) and SOCP (red), with SNR = 2/2 and 7/7 dB, N = 29, and $\Delta \phi = 0$}
\label{fig:sim_dr_nd}
\end{figure*}

From Fig. \ref{fig:sim_svd_rbgp_socp}, one can see that for the relatively high SNR case (10 dB and 3 dB), all methods can distinguish the two scatterers well when $\kappa = 1.2$. However, once they move close into one elevation resolution cell, SVD failed to detect double scatterers when $\kappa = 0.8, 0.4$ for both the low and high signal to noise ratio (SNR) conditions. In contrast, SOCP and RBGP can accurately estimate the position of double scatterers for all the cases, which exhibits the super-resolution power. If we further reduce $\kappa = 0.2$, SOCP and RBPG can distinguish the double scatterers for SNR = 10 dB. However, all methods failed to detect the double scatterers for SNR = 3 dB, which is not surprising according to the super-resolution power study reported in \cite{bib:Zhu2012b}. For the low SNR case (0 dB), note that even with $\kappa = 1.2$, SVD cannot distinguish the double scatterers. In contrast, both SOCP and RBPG can accurately estimate the position of double scatterers for $\kappa = 1.2, 0.8, 0.4$. In order to obtain more plausible evidence, a Monte Carlo simulation with 5,000 realizations per SNR value was performed to evaluate the detection rates of different normalized distances and schemes.

Fig. \ref{fig:sim_dr_nd} presents the detection rate $P_D$ as a function of normalized distance $\kappa$ at different SNR levels using SVD, RBPG and SOCP. The phase difference in this simulation, $\Delta \phi = 0$, is the worst case for detection \cite{bib:Zhu2012b}. The SNR of two sets of curves are 2 dB and 7 dB, respectively. The statistics confirm that the SVD approach does not have super-resolution power. For the poor SNR condition, the detection rate can not achieve $50\%$, even if the normalized distance is $\kappa = 1.2$. The RBPG approach behaves similarly to the SOCP approach by maintaining the super-resolution power of SL1MMER.

\subsection{Real Data}
\begin{figure*}
\centering
\subfloat[]{\includegraphics[width=0.454\textwidth]{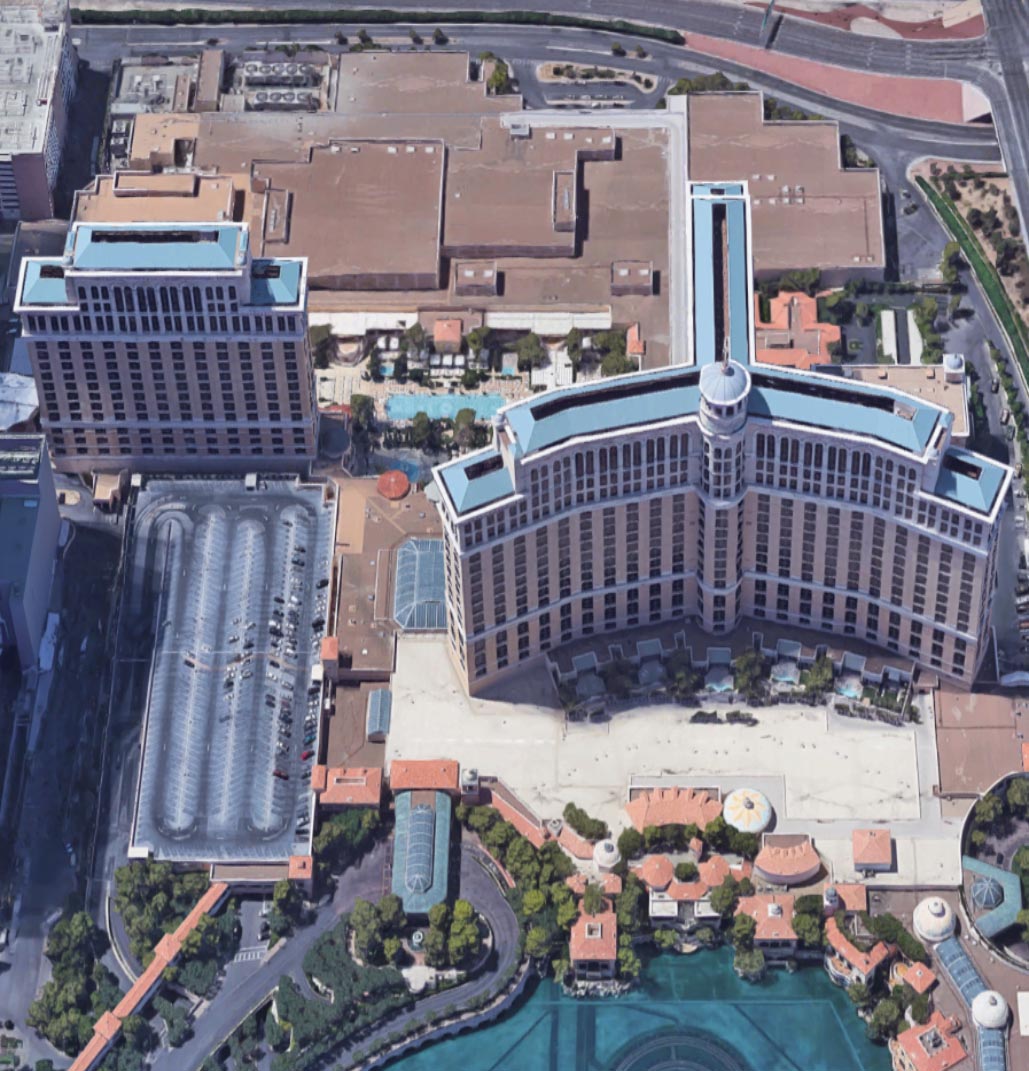}}
\qquad
\subfloat[]{\includegraphics[width=0.44\textwidth]{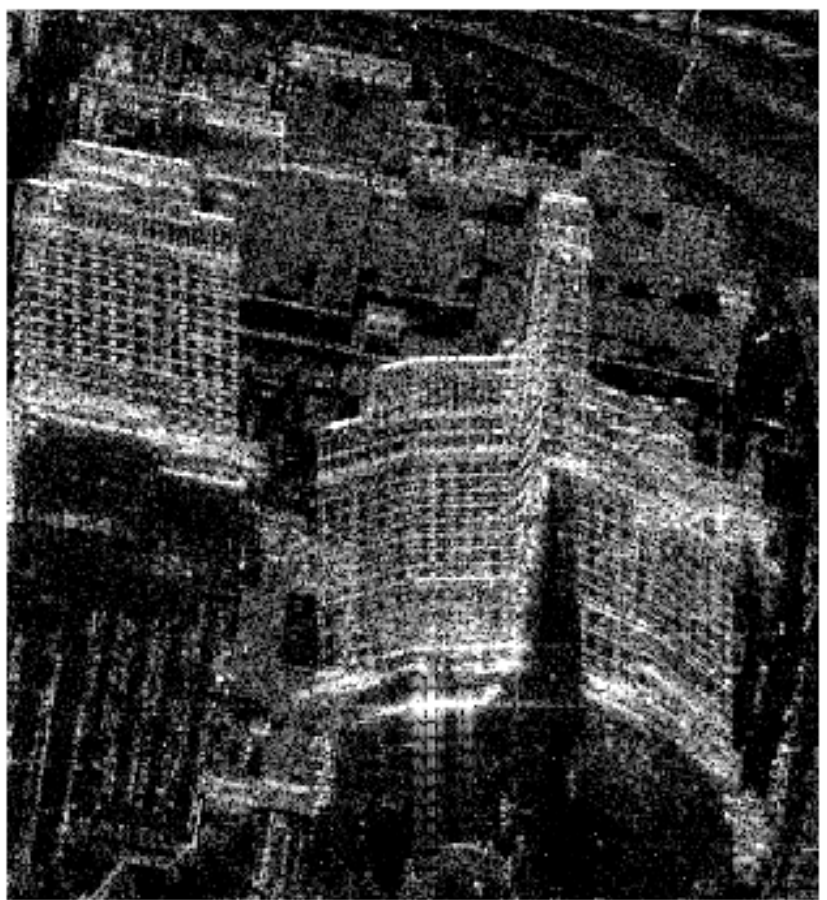}}
\caption{Test building: Bellagio hotel. (a) Optical image (Copyright Google) (b) TerraSAR-X mean intensity map.}
\label{fig:real_sar_optical}
\end{figure*}

For the real data experiment, we chose TerraSAR-X high resolution spotlight data with a slant-range resolution of 0.6 m and an azimuth resolution of 1.1 m. In order to be comparable to the results obtained with SOCP presented in \cite{bib:Zhu2012c}, we purposely used the same test data stack and test building. I.e., the stack taken over the city of Las Vegas consists of 29 images and has an elevation aperture size of about 269.5 m (i.e., the inherent elevation resolution is $\rho_s = 40 .5$ m, approximately a 20 m resolution in height with the elevation-to-height factor $\sin\theta$, where the incidence angle $\theta$ is $31.8^\circ$ here). The same test building -- Bellagio was chosen to demonstrate the SR power of the new approach, in comparision with the results shown in \cite{bib:Zhu2012c}, since its surrounding infrastructure exhibits strong scatterers that compete with the reflections from the building facade. Fig. \ref{fig:real_sar_optical} shows the optical image of the Bellagio from Google Maps and the TerraSAR-X mean intensity map.

\begin{figure*}
\centering
\subfloat[]{\includegraphics[width=0.454\textwidth]{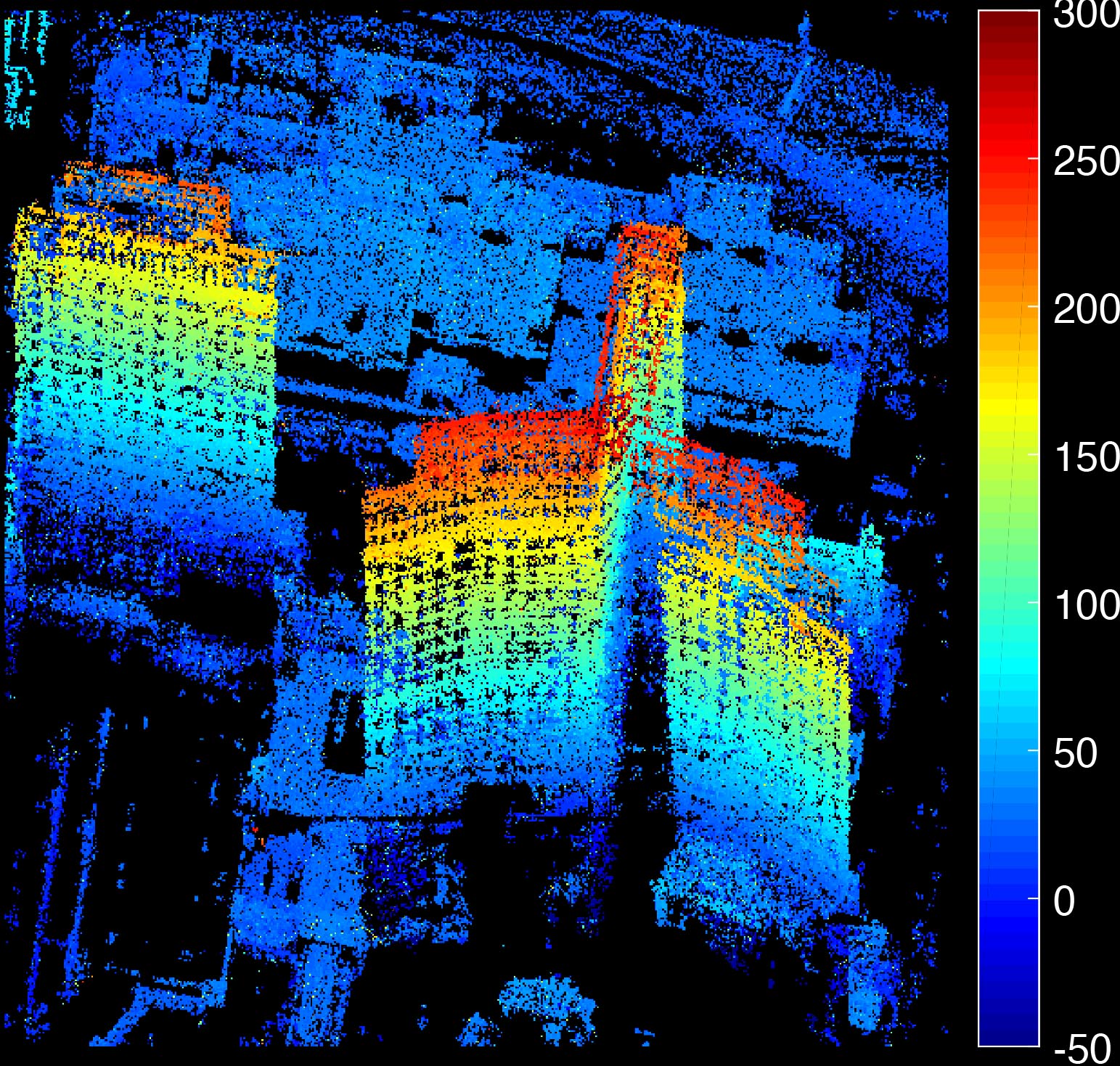}}
\qquad
\subfloat[]{\includegraphics[width=0.45\textwidth]{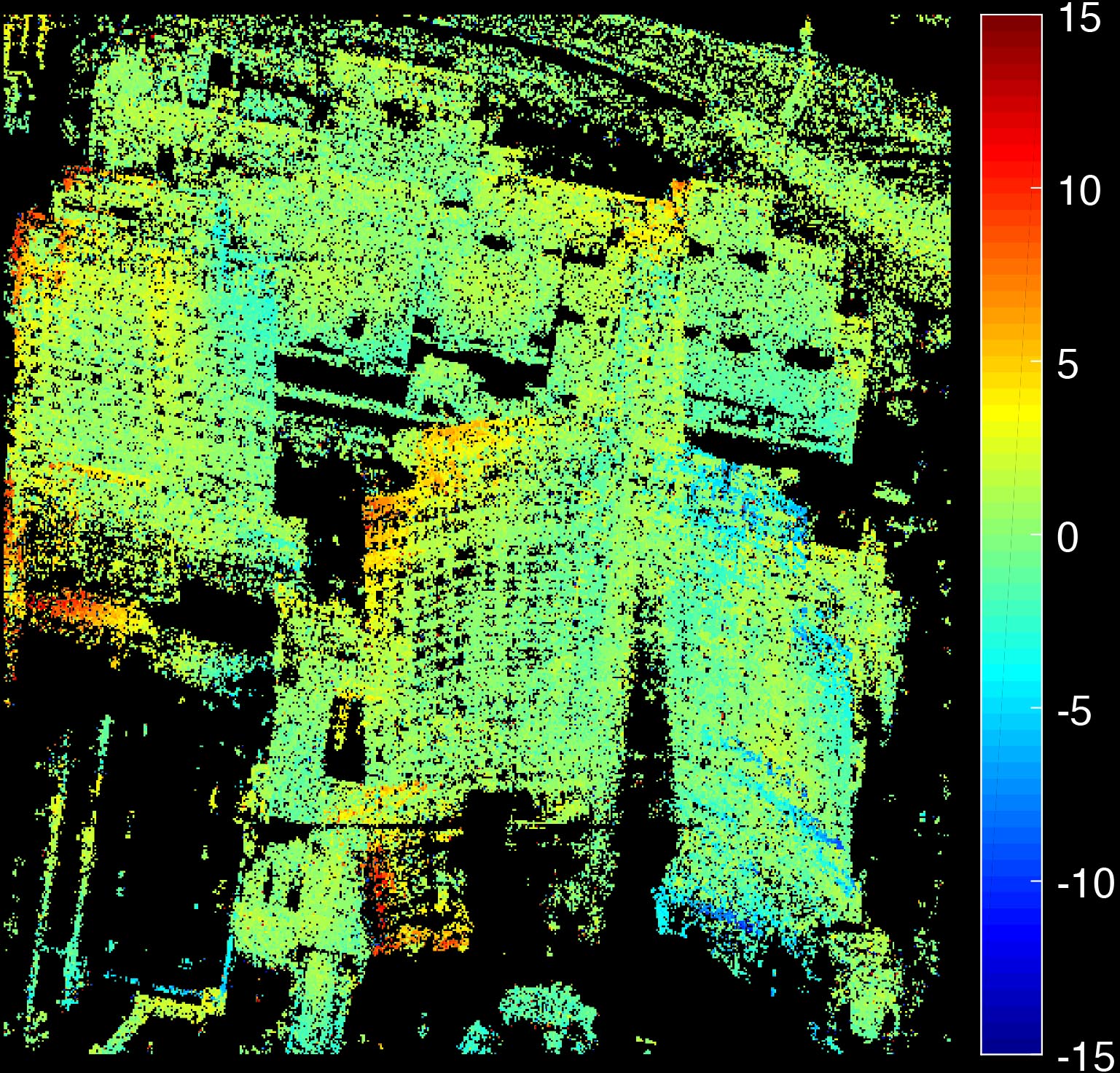}}
\caption{Test building: Bellagio hotel.  (a) Elevation in meters, estimated by RBPG  (b) Amplitude of seasonal motion estimates in millimeters, estimated by RBPG.}
\label{fig:real_elevation_motion}
\end{figure*}

Fig. \ref{fig:real_elevation_motion} (a) presents the fused topography estimates (i.e. the estimated elevations), of the detected single scatterers and double scatterers. The information increment contributed by the layover separation is significant, and the high density of detected double scatterers completes the structures of individual high-rise buildings. Fig. \ref{fig:real_elevation_motion} (b) shows the amplitude of the seasonal motion. The same as shown in \cite{bib:Zhu2012c}, the motion patterns are quite complex due to the fact that thermal dilation of buildings depends on many effects, like environmental air temperature, current sun illumination, internal cooling or heating, and the location of the major construction elements with respect to the facade. For the whole area, $29.1\%$ and $29.9\%$ of the scatterers detected by RBPG and SOCP, respectively, are found as double scatterers. From Tab. \ref{tab:ds_intersection}, we can see that $27.3\%$ of the double scatterers has been detected by both approaches.

\begin{table}
\begin{center}
\caption{Percentage of double scatterers detection for two approaches}
\label{tab:ds_intersection}
\begin{tabular}{ccc}
\toprule
DS only detected by RBPG & Intersection & DS only detected by SOCP\\
\midrule
1.8\% & 27.3\% & 2.6\%\\
\bottomrule
\end{tabular}
\end{center}
\end{table}

\begin{figure*}
\centering
\subfloat[]{\includegraphics[width=0.454\textwidth]{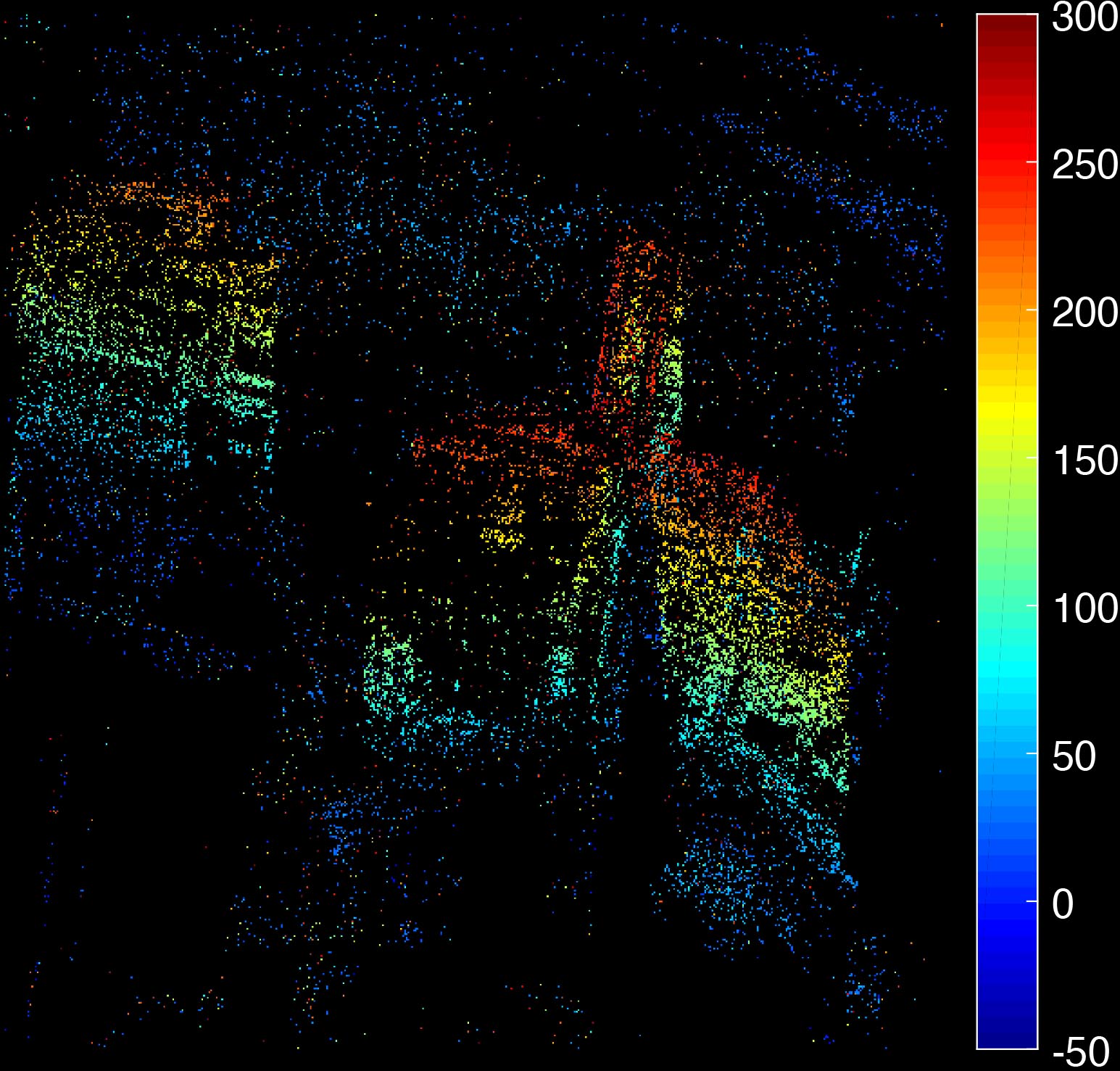}}
\qquad
\subfloat[]{\includegraphics[width=0.451\textwidth]{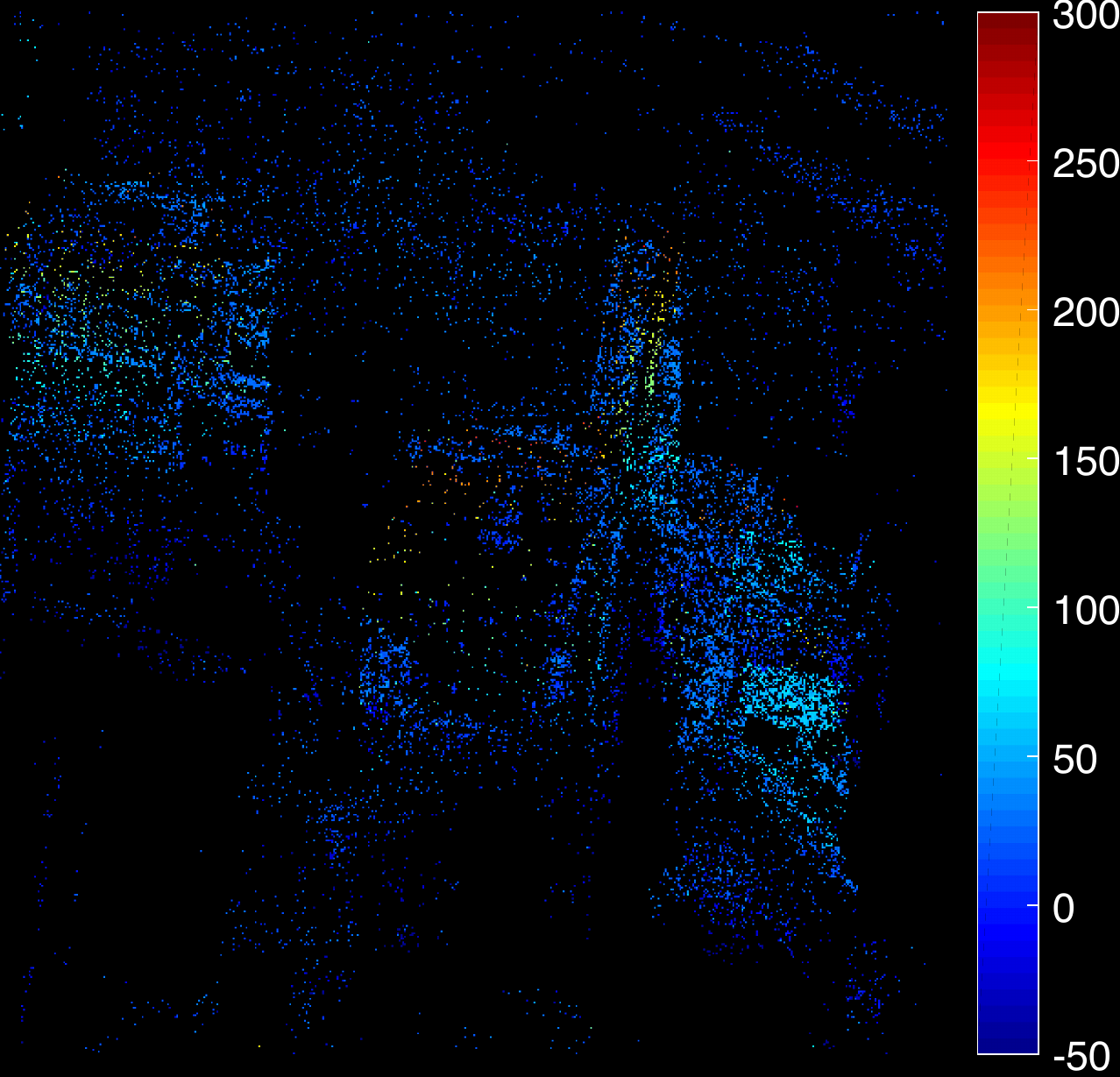}}
\caption{Test building: the Bellagio. (a) Top layer, mainly caused by returns from building facade (b) Ground layer, mainly caused by returns from ground structures.}
\label{fig:real_top_ground}
\end{figure*}

Fig. \ref{fig:real_top_ground} presents the estimated elevation of the two layers of the detected double scatterers, with the two layers consisting of a top layer mainly caused by the reflections from the building facade and a ground layer caused by reflections from lower buildings or ground infrastructures. The gradation of elevation estimates on the top layer [see Fig. \ref{fig:real_top_ground} (a)] and the homogeneity in the ground layer [see Fig. \ref{fig:real_top_ground} (b)] suggest the correctness of the elevation estimation and layover separation capability. Comparable to SOCP \cite{bib:Zhu2012c}, the full structure of the high-rise building is almost captured with only the detected double scatterers.

\subsection{Large-Scale Demonstration}
To validate our approach, we chose a large-scale test area covering the whole city of Munich. The TerraSAR-X data stack is composed of 78 very high-resolution spotlight images and covers approximately 50 $\mathrm{km}^2$. Four-dimensional point clouds with a density of about one million points per square kilometer are reconstructed. The experiments were carried out on a high-performance computer at Lebnitz-Rechnung-Zentrum (LRZ) with about 2,000 cores. With the same number of cores, the run time using the SOCP approach is estimated to be 120 CPU hours whereas RBPG took only six CPU hours. The new approach speeds up by a factor of 20 for this large-scale case.

\begin{figure}
\centering
\includegraphics[width=0.497\textwidth]{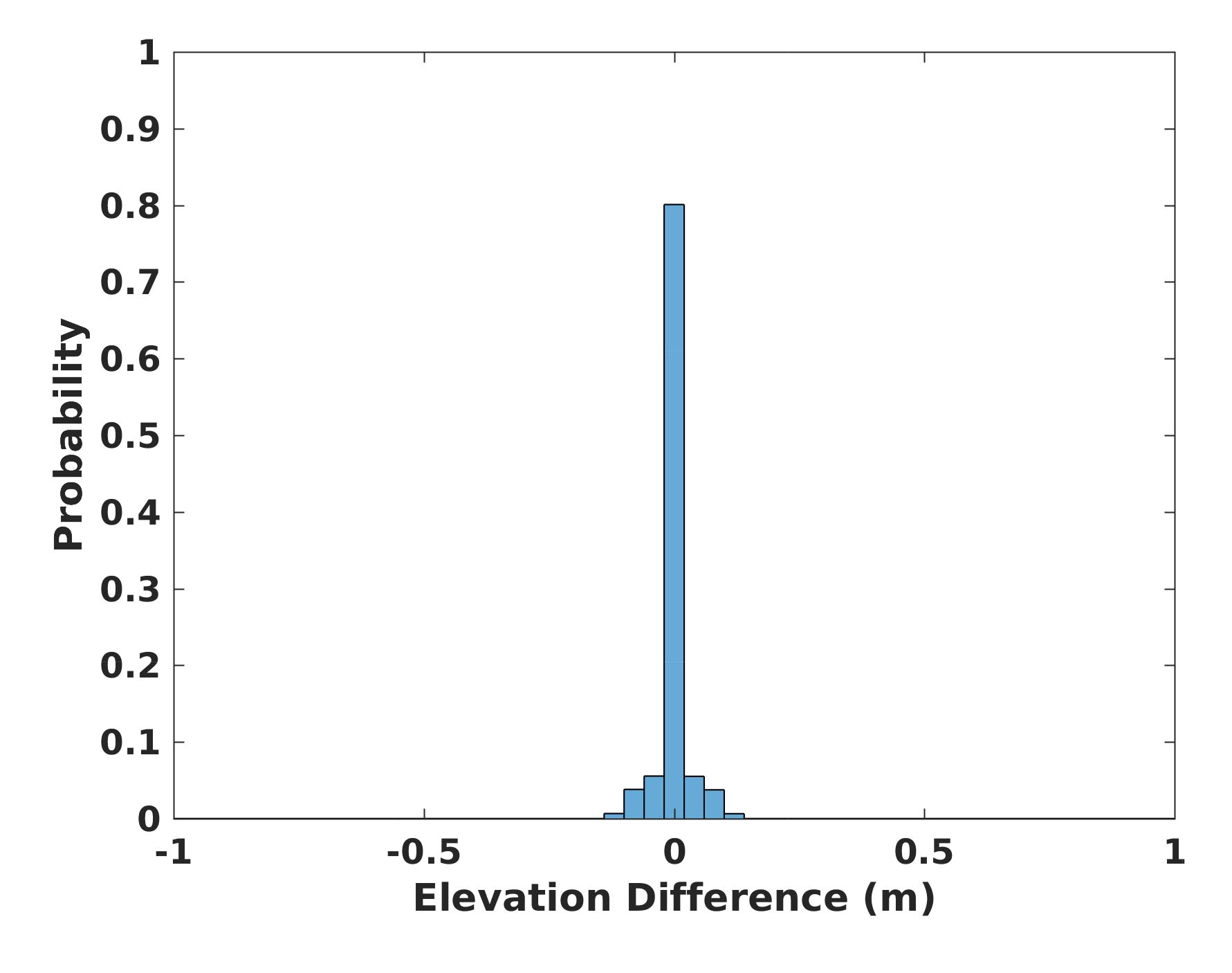}
\caption{Histogram of elevation differences of both methods for large-scale demonstration.}
\label{fig:large_scale_hist}
\end{figure}

The histogram of elevation differences of both methods is shown in Fig. (\ref{fig:large_scale_hist}). Note that most of the elevation differences collapse at zero, which indicates the estimation accuracy of RBPG as being similar to SOCP.

\begin{figure*}
\centering
\subfloat[]{\includegraphics[width=0.8\textwidth]{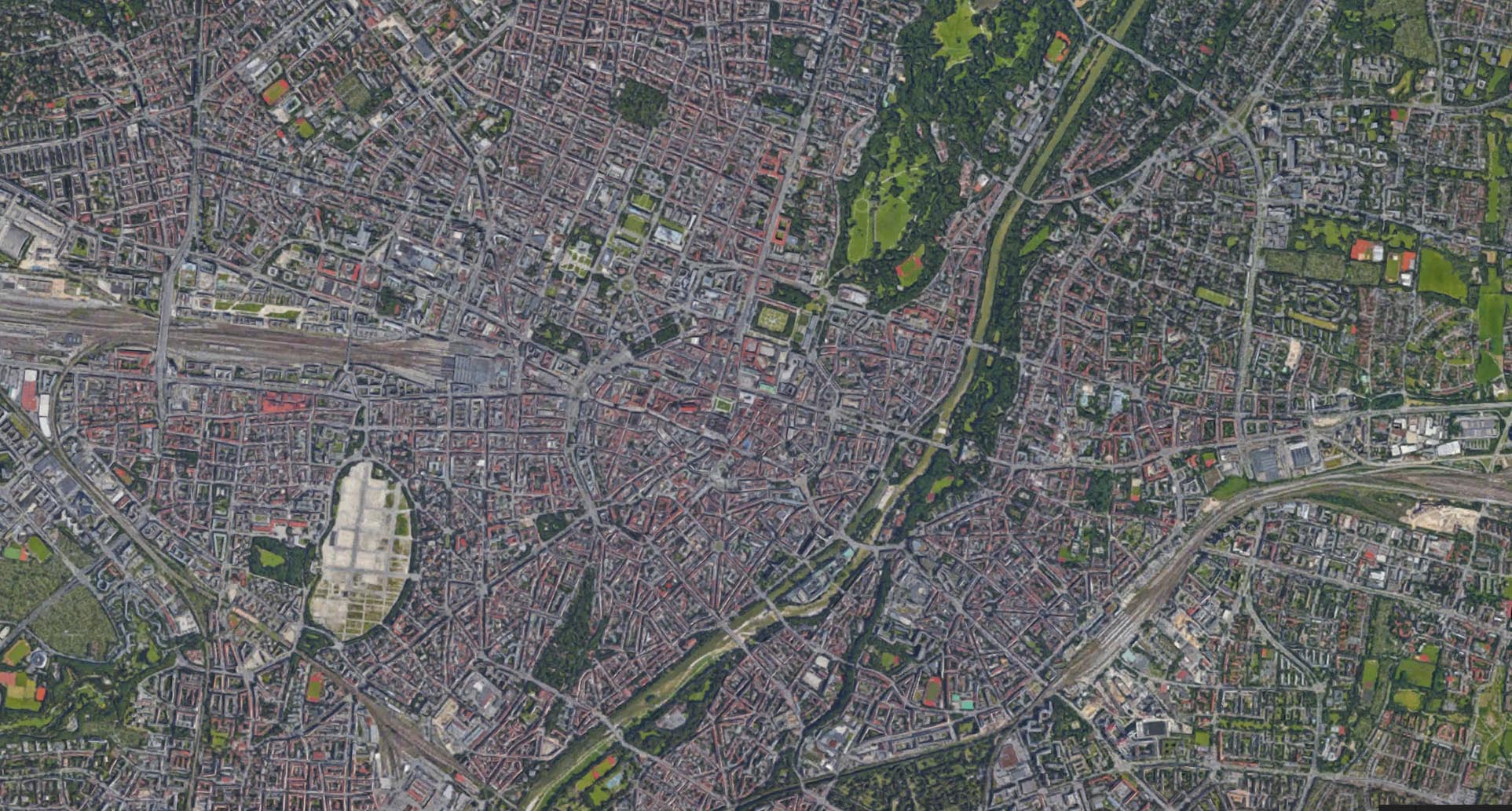}}
\hfil
\subfloat[]{\includegraphics[width=0.8\textwidth]{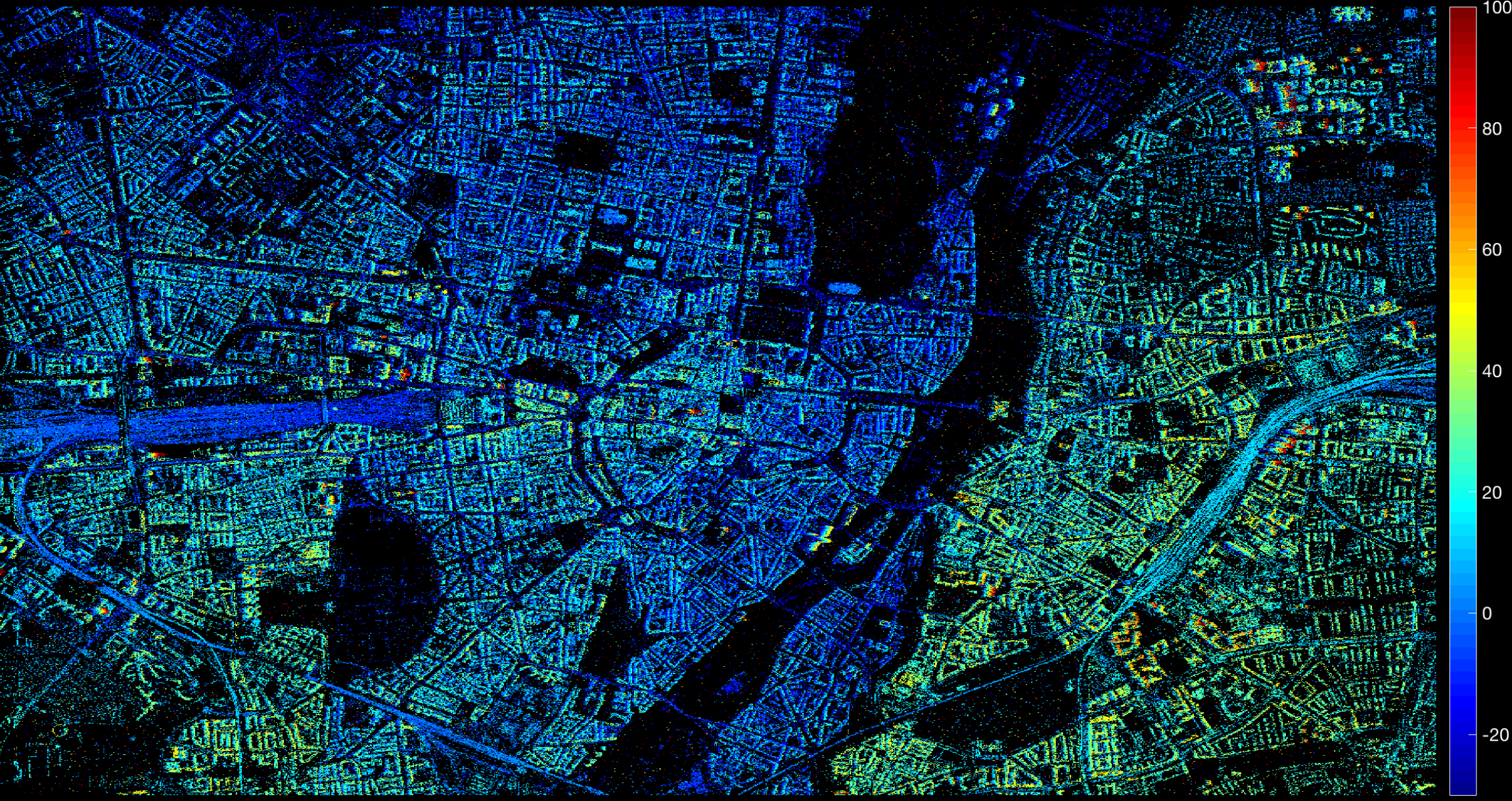}}
\caption{Test Area: Munich. (a) optical image of the test area \textcopyright Google. (b) reconstructed point cloud with height color-coded. }
\label{fig:large_scale_demo}
\end{figure*}

Fig. \ref{fig:large_scale_demo} (b) shows the elevation estimates of InSAR stacks. As a comparison, we show the corresponding area of the optical image in Fig. \ref{fig:large_scale_demo} (a).

\begin{figure*}
\centering
\subfloat[]{\includegraphics[width=0.46\textwidth]{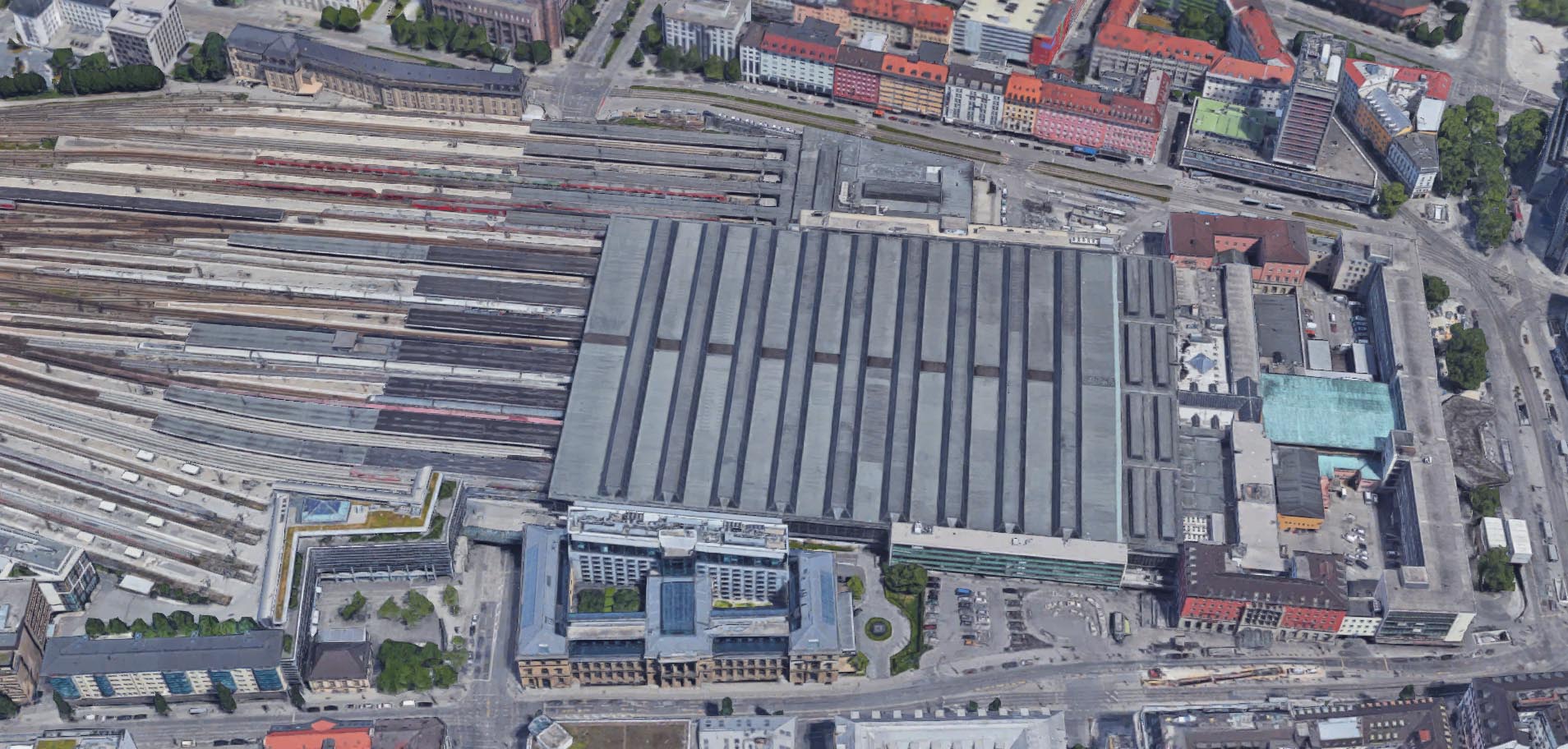}}
\hfil
\subfloat[]{\includegraphics[width=0.44\textwidth]{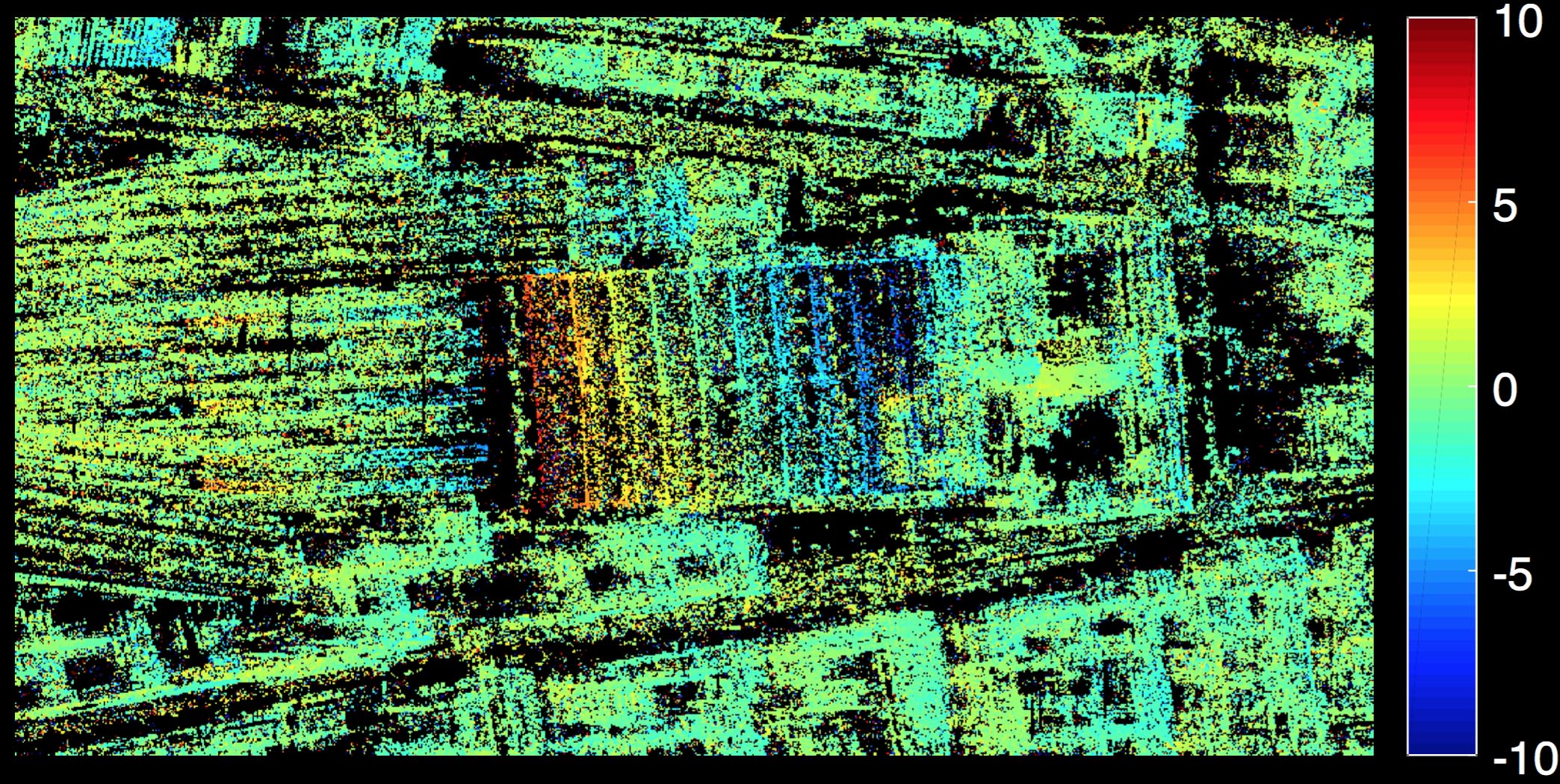}}
\caption{Estimated seasonal deformation of the central train station in Munich. Red color indicates movement toward the sensor, blue color away from the sensor.}
\label{fig:seasonal_defo_demo}
\end{figure*}

A clear seasonal deformation is observed in the central train station in Munich, which is caused by thermal dilation of the metallic building structure. As one can see in Fig. \ref{fig:seasonal_defo_demo}, a red color indicates movement toward the sensor, a blue color means movement away from the sensor, with the amplitude of the deformation is up to 10 mm/year.

\begin{figure*}
\centering
\subfloat[]{\includegraphics[width=0.44\textwidth]{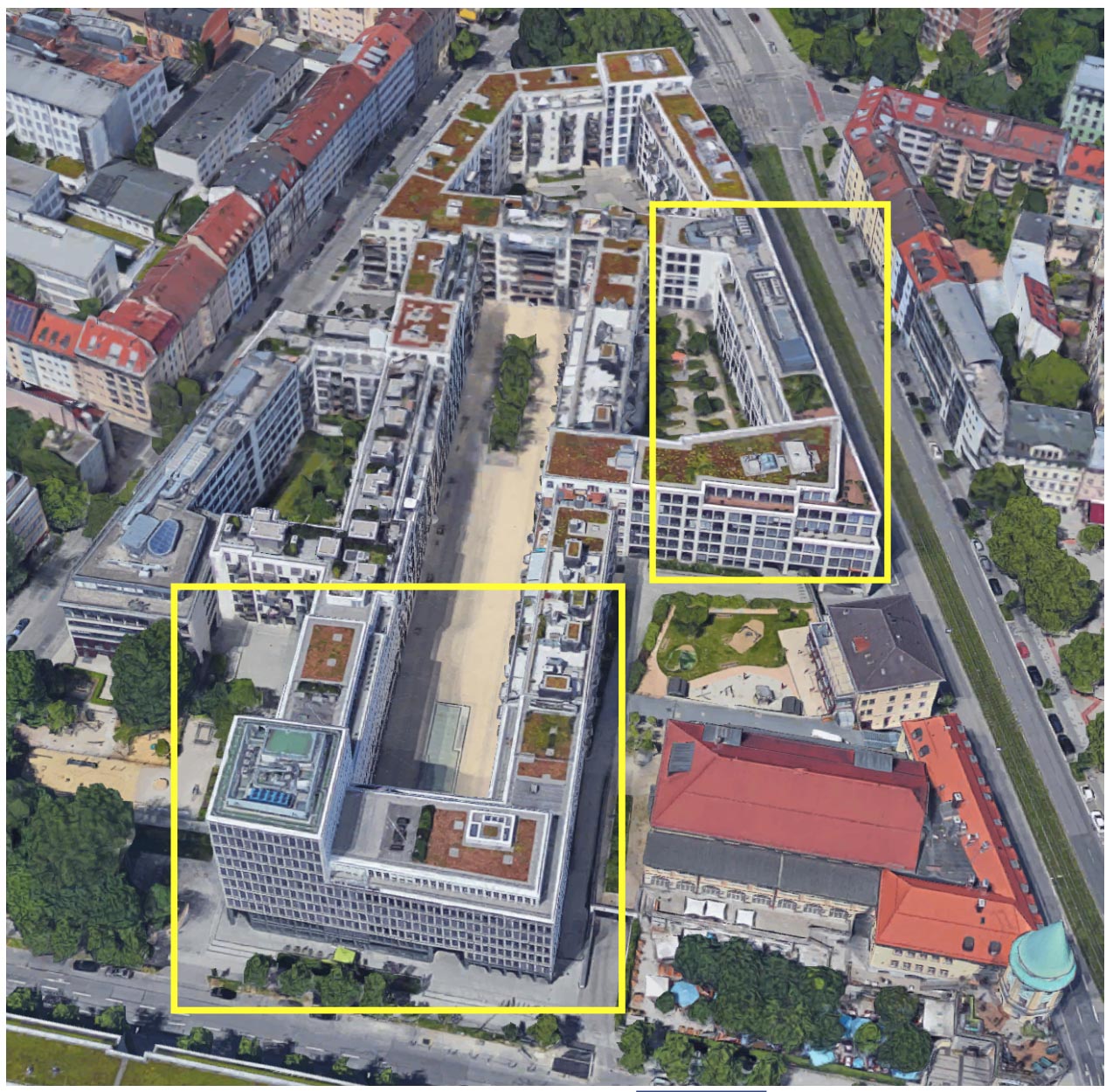}}
\hfil
\subfloat[]{\includegraphics[width=0.52\textwidth]{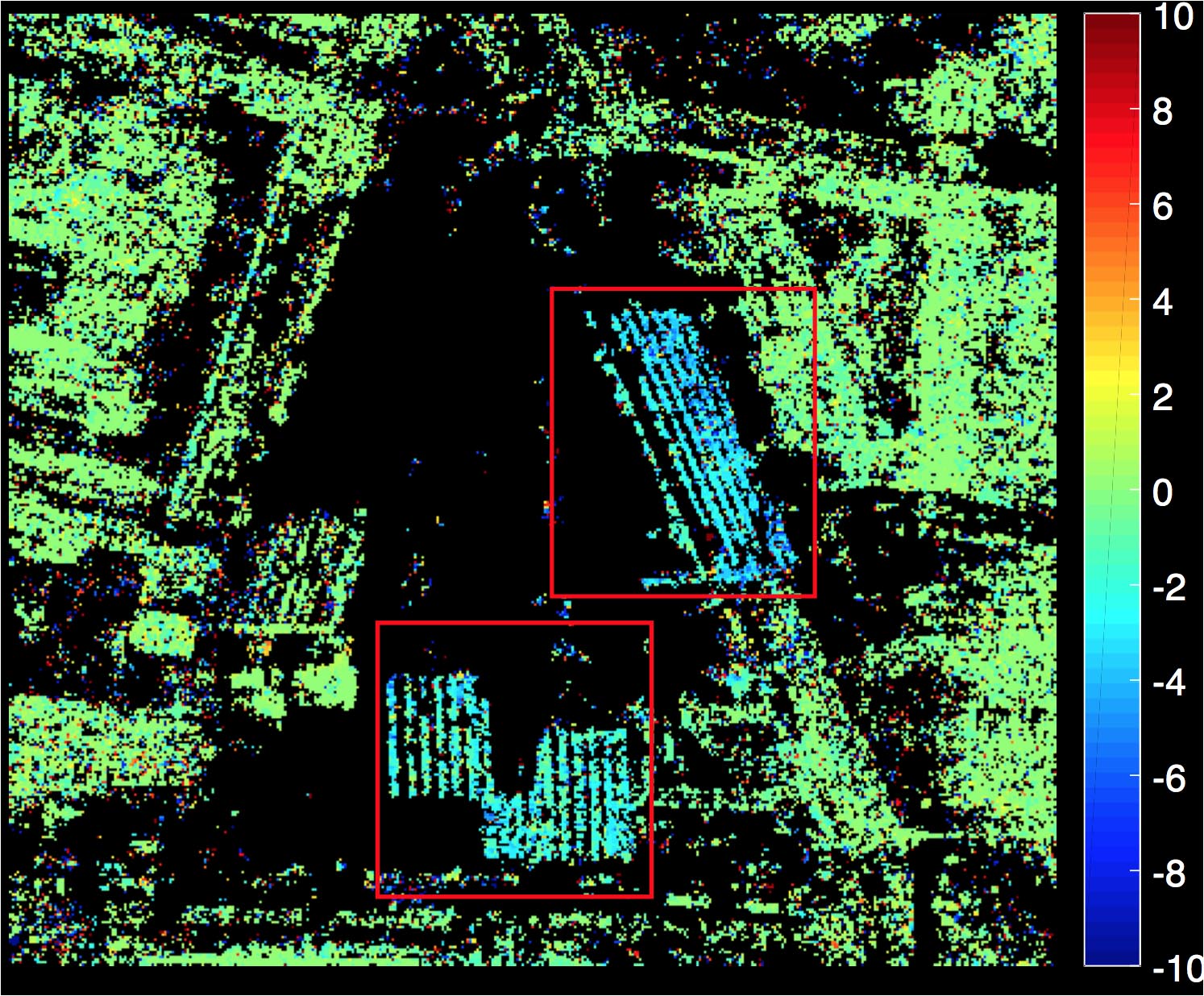}}
\vfil
\subfloat[]{\includegraphics[width=\textwidth]{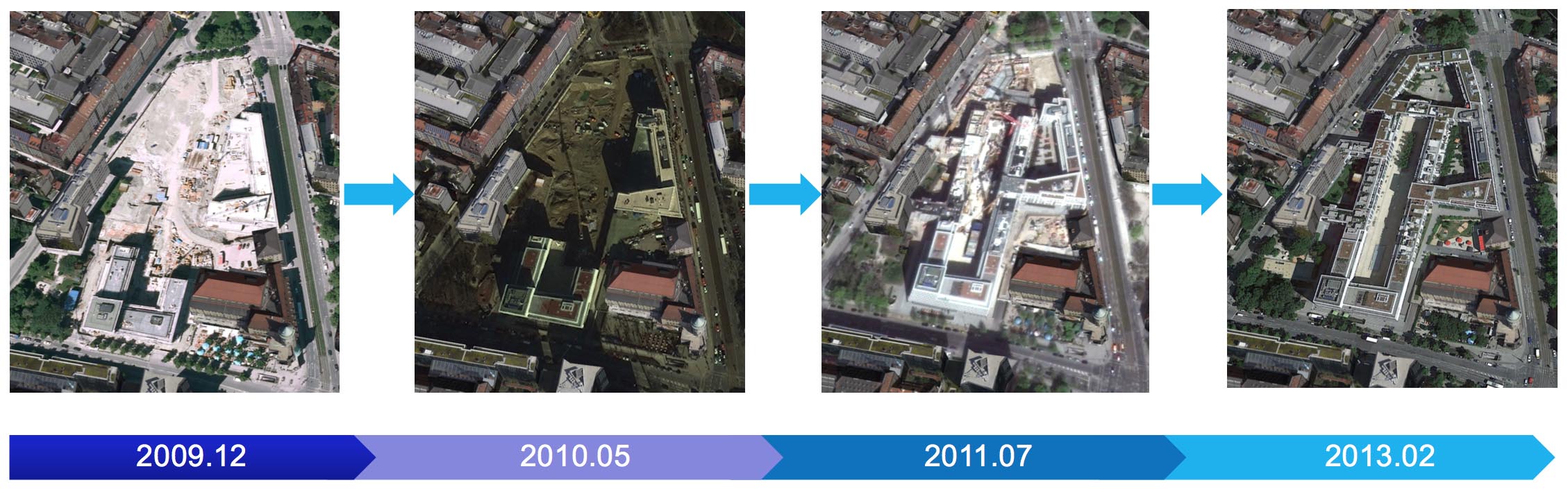}}
\caption{(a) Linear deformation caused by the construction of new buildings. (b) Time-lapse of optical images from Google Earth.}
\label{fig:linear_defo_demo}
\end{figure*}

An another interesting example shows an area near the LowenBrau Keller, which is a famous beer company. We chose this area due to the clear linear deformation patterns. From the corresponding time-lapse of optical images from Google Earth illustrated in Fig. \ref{fig:linear_defo_demo}, we can see that the linear deformation is caused by the construction of new buildings.

\section{Conclusion}
In this work, we have proposed a fast and accurate optimization approach for solving complex-valued $L_1$ regularized least squares -- a widely employed optimization formulation in radar remote sensing. Tomographic SAR processing was used as a practical application. Experiments using simulated data and real data demonstrate that the new approach retains the super-resolution power of second order sparse recovery in TomoSAR processing and speeds it up for one or two orders of magnitude, which allows for the operational processing of large-scale problems. Combining the proposed optimization approach with the processing strategy proposed in \cite{bib:Wang2014}, a further speed-up of about 50 times can be expected. While our exposition uses TomoSAR in remote sensing as an example, the proposed algorithm can be generally used for spectral estimation.


\vfill

\begin{IEEEbiography}[{\includegraphics[width=1in,height=1.5in,clip,keepaspectratio]{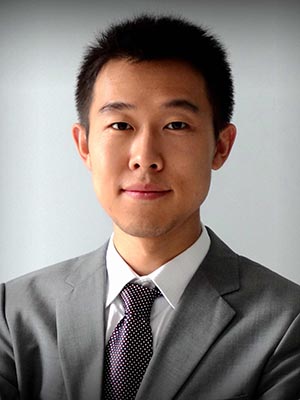}}]{Yilei Shi}
received his Diploma degree in Mechanical Engineering from Technical University of Munich (TUM), Germany, in 2010. He is currently a research associate with the Chair of Remote Sensing Technology, Technical University of Munich.

His research interests include fast solver and parallel computing for large-scale problems, advanced methods on SAR and InSAR processing, machine learning and deep learning for variety data sources, such as SAR, optical images, medical images and so on; PDE related numerical modeling and computing.
\end{IEEEbiography}

\begin{IEEEbiography}[{\includegraphics[width=1in,height=1.5in,clip,keepaspectratio]{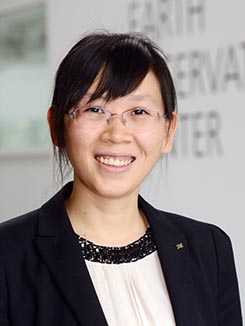}}]{Xiao Xiang Zhu}
(S'10--M'12--SM'14) received the Master (M.Sc.) degree, her doctor of engineering (Dr.-Ing.) degree and her “Habilitation” in the field of signal processing from Technical University of Munich (TUM), Munich, Germany, in 2008, 2011 and 2013, respectively.

She is currently the Professor for Signal Processing in Earth Observation (www.sipeo.bgu.tum.de) at Technical University of Munich (TUM) and German Aerospace Center (DLR); the head of the department of EO Data Science at DLR; and the head of the Helmholtz Young Investigator Group ”SiPEO” at DLR and TUM. Prof. Zhu was a guest scientist or visiting professor at the Italian National Research Council (CNR-IREA), Naples, Italy, Fudan University, Shanghai, China, the University  of Tokyo, Tokyo, Japan and University of California, Los Angeles, United States in 2009, 2014, 2015 and 2016, respectively. Her main research interests are
remote sensing and earth observation, signal processing, machine learning and data science, with a special application focus on global urban mapping.

Dr. Zhu is a member of young academy (Junge Akademie/Junges Kolleg) at the Berlin-Brandenburg Academy of Sciences and Humanities and the German National  Academy of Sciences Leopoldina and the Bavarian Academy of Sciences and Humanities. She is an associate Editor of IEEE Transactions on Geoscience and Remote Sensing.
\end{IEEEbiography}

\begin{IEEEbiography}[{\includegraphics[width=1in,height=1.5in,clip,keepaspectratio]{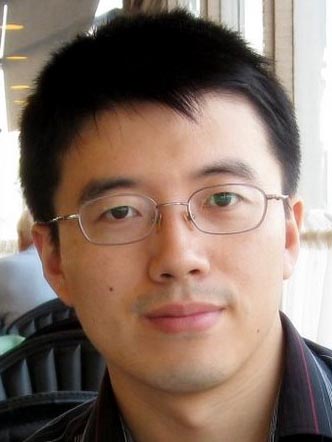}}]{Wotao Yin}
received his Ph.D. degree in operations research from Columbia University, New York, NY, USA, in 2006, respectively. He is currently a Professor with the Department of Mathematics, University of California, Los Angeles, CA, USA. His research interests include computational optimization and its applications in signal processing, machine learning, and other data science problems.

During 2006--2013, he was at Rice University. He was the NSF CAREER award in 2008, the Alfred P. Sloan Research Fellowship in 2009, and the Morningside Gold Medal in 2016, and has coauthored five papers receiving best paper-type awards. He invented fast algorithms for sparse optimization and has been leading the research of optimization algorithms for large-scale problems.
\end{IEEEbiography}

\begin{IEEEbiography}[{\includegraphics[width=1in,height=1.5in,clip,keepaspectratio]{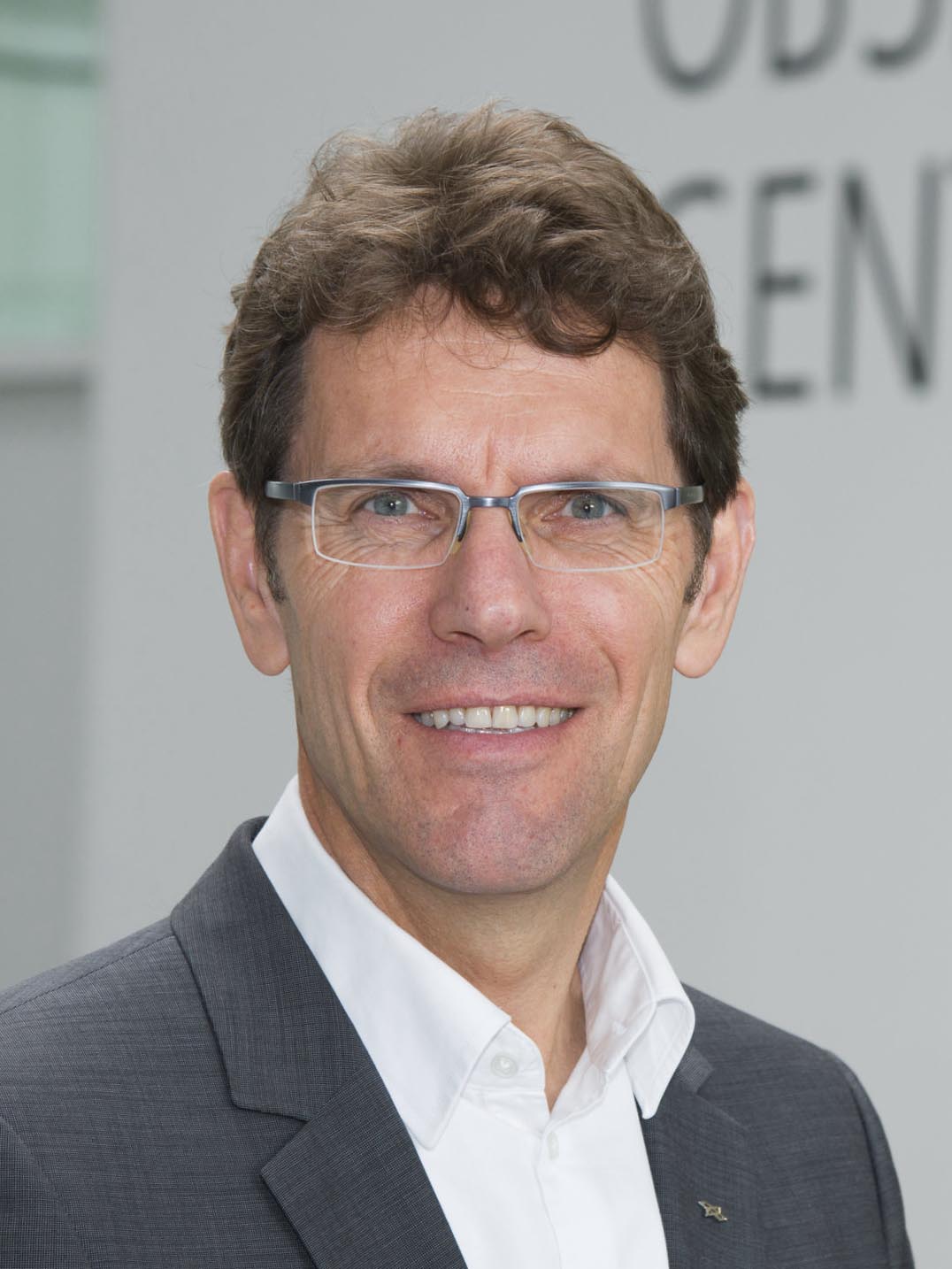}}]{Richard Bamler}
(M'95--SM'00--F'05) received his Diploma degree in Electrical Engineering, his Doctorate in Engineering, and his “Habilitation” in the field of signal and systems theory in 1980, 1986, and 1988, respectively, from the Technical University of Munich, Germany.

He worked at the university from 1981 to 1989 on optical signal processing, holography, wave propagation, and tomography. He joined the German Aerospace Center (DLR), Oberpfaffenhofen, in 1989, where he is currently the Director of the Remote Sensing Technology Institute.

In early 1994, Richard Bamler was a visiting scientist at Jet Propulsion Laboratory (JPL) in preparation of the SIC-C/X-SAR missions, and in 1996 he was guest professor at the University of Innsbruck. Since 2003 he has held a full professorship in remote sensing technology at the Technical University of Munich as a double appointment with his DLR position. His teaching activities include university lectures and courses on signal processing, estimation theory, and SAR. Since he joined DLR Richard Bamler, his team, and his institute have been working on SAR and optical remote sensing, image analysis and understanding, stereo reconstruction, computer vision, ocean color, passive and active atmospheric sounding, and laboratory spectrometry. They were and are responsible for the development of the operational processors for SIR-C/X-SAR, SRTM, TerraSAR-X, TanDEM-X, Tandem-L, ERS-2/GOME, ENVISAT/SCIAMACHY, MetOp/GOME-2, Sentinel-5P, Sentinel-4, DESIS, EnMAP, etc.

Richard Bamler’s research interests are in algorithms for optimum information extraction from remote sensing data with emphasis on SAR. This involves new estimation algorithms, like sparse reconstruction, compressive sensing and deep learning.
\end{IEEEbiography}

\end{document}